\documentclass{emulateapj}
\usepackage{natbib}
\usepackage{epsfig}
\usepackage{graphicx} 
\usepackage{subfigure}
\usepackage{epstopdf}
\usepackage{float}
\usepackage{amsmath}
\usepackage{color}
\usepackage{amssymb}
\usepackage{amsfonts}
\usepackage{units}
\usepackage{mathtools}
\usepackage{bm}
\usepackage[colorlinks,linkcolor=blue,anchorcolor=green,citecolor=blue]{hyperref}
\renewcommand{\d}{\mathrm{d}}

\newcommand{\bea}{\begin{eqnarray}}
\newcommand{\eea}{\end{eqnarray}}
\newcommand{\be}{\begin{equation}}
\newcommand{\ee}{\end{equation}}
\newcommand{\rund}[1]{\left(#1\right)}

\def\elabel#1{\label{eq:#1}}
\renewcommand{\exp}{\mathrm{exp}}

\shorttitle{Lensing and FRB DMs}
\shortauthors{Er et al.} 
\begin{document}

\title{The effects of plasma lensing on the inferred dispersion measures of Fast Radio Bursts}

\author{Xinzhong Er\altaffilmark{1}, Yuan-Pei Yang\altaffilmark{1}, Adam Rogers\altaffilmark{2}} 

\affil{$^1$South-Western Institute for Astronomy Research, Yunnan University, Kunming, Yunnan, P.R.China; xer@ynu.edu.cn;\\
$^2$Department of Physics and Astronomy, University of Manitoba, Winnipeg, MB, R3T 2N2, Canada}

\begin{abstract}
  Radio signals are delayed when propagating through plasma. This type
  of delay is frequency-dependent and is usually used for estimating
  the projected number density of electrons along the line of sight,
  called the dispersion measure. The dense and clumpy distribution of
  plasma can cause refractive deflections of radio signals, analogous
  to lensing effects. Similar to gravitational lensing, there are two
  contributions to the time delay effect in plasma lensing: a
  geometric delay, due to increased path length of the signal, and a
  dispersive delay due to the change of speed of light in a plasma
  medium. We show the delay time for two models of the plasma
  distribution, and point out that the estimated dispersion measure
  can be biased. Since the contribution of the geometric effect can be
  comparable to that of the dispersive delay, the bias in the measured
  dispersion measure can be dramatically large if plasma lensing
  lensing effects are not taken into account when signals propagate
  through a high-density gradient clump of plasma.
\end{abstract}


\section{Introduction}

Fast radio bursts (FRBs) are a new kind of radio transient with millisecond duration. These mysterious events are characterized by an excess dispersion measure (DM) with respect to the Galactic contribution as well as high brightness. Evidence is emerging that FRBs are distributed isotropically on the sky \citep[e.g.][]{thornton2013,Shannon2018,frbreview}, however the physical origin of these bursts is still unknown. At the time of writing, the total number of the published FRBs is around 100 (see FRB Catalog\footnote{http://frbcat.org/} of \citet{petroff2016}). Several of these bursts show repeating behaviors, including FRB 121102 \citep{repeatingFRB,extragalacticFRB} and FRB 180814 \citep{chime2019a}, and eight FRBs recently discovered by the Canadian Hydrogen Intensity Mapping Experiment \citep[CHIME;][]{2019arXiv190803507T,cas19}. Thanks to interferometric localizations, FRB 121102, FRB 180924, FRB 181112 and FRB 190523 have been localized to sufficient accuracy to identify their host galaxies \citep{extragalacticFRB,ban19,rav19,pro19}. The first repeating burst, FRB 121102, has been found to be located in a star-forming dwarf galaxy at $z = 0.19273$ and associated with a persistent radio source \citep{extragalacticFRB,tendulkar2017,marcote2017}. FRB 180924 is localized to a position $4~{\rm kpc}$ from the center of a luminous galaxy at redshift $z=0.3214$ \citep{ban19}. FRB 190523 is found to be associated with a massive galaxy with low specific star-formation rate at a redshift of $z=0.66$ \citep{rav19}. Since DM and redshift have been measured for these FRBs, they can be used as an intergalactic and cosmological probe \citep[e.g.][]{den14,yan16,yan17,li19}.

As with any radio transients at cosmological distance, FRBs are dispersed when they propagate in ionized gas, i.e. free electrons. Generally, the lower the frequency of a signal, the longer the delay time, which has been found in all FRB observations. To describe this behaviour, let us define $\Delta t$ as the delay time\footnote{In gravitational lensing, the time delay sometimes refers to the difference in the arrival times between multiple lensed images. In this work, the time delay represents the delay in excess of the unlensed case. This is equivalent to measuring time delay with respect to the arrival time of the signal in the high frequency limit.} of a burst at frequency $\nu$. The delay time -- frequency relation of FRBs is nearly consistent with the classical dispersion of an electromagnetic wave in cold plasma \citep{lorimer2007}, $\Delta t\propto{\rm DM}\nu^{-2}$, where ${\rm DM}$ directly reflects the free electron column density along the line of sight, e.g., ${\rm DM}\equiv\int n_edl$. A slight discrepancy from the quadratic time delay relationship has been found through observations \citep[e.g.][]{2013Sci...341...53T,2016MPLA...3130013K}, suggesting the plasma may also be emitting \citep[e.g.][]{frbreview}. For FRBs, one of the most important features is the excess DM with respect to the Milky Way's contribution, suggesting that the bursts have a cosmological origin \citep{thornton2013}.

Generally, since FRBs seem to be isotropically distributed over the
sky, we expect refractive lensing to occur due to a chance alignment
of an FRB source with a foreground lens object. The lens may be a
small plasma inhomogeneity within the host galaxy of the FRB
\citep{FRBplasma1}, an intervening galaxy or cluster halo may act as a
strong gravitational lens \citep{li14,dai17,li18}, or an isolated and
extragalactic compact object may act as a gravitational microlens
\citep{zhe14,mun16}. In particular, \cite{pro19} recently found that
FRB 181112 passed through a foreground galaxy halo, and they proved
that the burst observation characteristics can be used to constrain
the plasma properties (e.g., magnetic field and turbulence). In
addition, the probability of a radio signal propagating through a clump of
plasma within the Milky Way is also high \citep{2002astro.ph..7156C}. We
therefore consider three possible locations for plasma lenses: within
the Milky Way, an intervening galaxy or galaxy cluster, and within the
host galaxy. In the case of an intervening galaxy, we assume the
galaxy is not sufficiently aligned with the line of sight to
substantially contribute any gravitational lensing effect, but that a
plasma inhomogeneity within the galaxy is sufficiently positioned to
act as a lens to the distant source.

Plasma lensing plays an important role in the ``Extreme Scattering
Events'' (ESEs) that are seen in the light curves of some active
galactic nuclei (AGN) and pulsars. Plasma lensing is the phenomenon of
radiation travelling along deflected paths due to the variable
electron density across the plane of the sky
\citep[e.g.][]{1957PhT....10h..30R,ESEIntro1}. These ESEs are
consistent with plasma lensing from $\sim$Astronomical Unit (AU)
structures in the Milky Way
\citep[e.g.][]{ESE0,2012MNRAS.421L.132P,ESE1,2014MNRAS.442.3338P,2015ApJ...808..113C}. Several
models of plasma lenses have been proposed from analytical
distributions or by fitting the observations
\citep[e.g.][]{cleggFey,1990ARA&A..28..561R,romani87,ESE3,ESE4,2016MNRAS.458.1289L,ErRogers18},
and even include magnetic fields \citep[e.g.][]{2019MNRAS.484.5723L}.
It has been also suggested that pulsar scintillation is caused by
scattering due to plasma structures
\citep[e.g.][]{stinebring,2006ApJ...637..346C,coles2010,2016ApJ...817...16C,nanoGravPSR,ESE5,kerr2018,2018arXiv181007231S,2019MNRAS.486.2809G}.
Recently, \citet{FRBplasma1} proposed that the amplitude of an FRB can
be strongly modulated by plasma lenses in the host galaxy. The
complex properties of plasma lenses might account for the
observations of repeating FRBs. For example, strong focusing by
plasma lenses can produce large intensity variations with factors of
10-100, which might account for the intermittency seen from FRB 121102
\citep{FRBplasma1,hessels2018}. If a plasma lens acts on an FRB that
is observed as a repeater, the properties of the plasma lens,
including lens size, density and transverse velocity, can be
constrained by the observation of the DM variation of the repeating
FRBs \citep{yang2017}.

Similar to gravitational lensing \citep{SEF}, plasma lensing also
causes time delays. In contrast to gravitational lensing, plasma
lensing leads to different observable phenomena. For instance, suppose
the deflection caused by plasma lensing of a background source is
small. Then the image separations between the multiple images (if
multiple imaging occurs at all), are extremely difficult to
resolve. In this case, the time delay between images in a lensing
system are unlikely to be directly measured due to their tiny angular
separation. However, in plasma lensing, the time delay is
frequency-dependent, offering an entirely unique avenue to study the
structure of the lens and the source. Moreover, plasma lenses are very
versatile in terms of their magnification properties. Generally plasma
acts like a diverging lens responsible for de-magnification of
background sources. However, based on the particular geometry of lens,
source and observer, can also cause substantial magnification to occur
\citep[e.g.][]{kerr2018,ErRogers18,2018MNRAS.481.2685D}. Plasma lenses
that are under-dense compared to the surrounding interstellar medium
(ISM) behave like converging lenses and magnify background sources
\citep[e.g.][]{2012MNRAS.421L.132P}.

The frequency-dependent delay caused by plasma lenses shows behaviour
that is distinct from the classical dispersion relation, since the
change of path length causes an extra delay in the propagation of the
signal, i.e. a geometric delay. As we will see in the next sections,
the geometric delay is proportional to the square of the deflection
angle, and thus has a dependence on wavelength to the fourth
power. When the gradient of the plasma density is large, the geometric
term can dominate the total delay time.  For instance, the
2-dimensional dynamic power spectra of some pulsars show organized
parabolic structures, which suggests significant geometric
contributions to the time delay
\citep{stinebring,2007ASPC..365..254S}. Moreover, it has been noted
that the DM variations with frequency can be used to study
sub-structures in the ionized ISM \citep[e.g.][and references
  therein]{2016ApJ...817...16C, 2019A&A...624A..22D,
  2019arXiv190300426L}. Therefore, if an FRB passes through such an
ionized sub-structure in the ISM, the frequency-dependent delay time
of plasma lensing would affect the observed apparent dispersion
relation of the FRB. Plasma lensing can also induce other
frequency-dependent effects, such as displacement, magnification
(scintillation), and distortion in the multiple images of a background
source.

In this work, we focus on the delay time of FRBs induced by various
plasma lens models, including exponential and power-law models. In
addition, we consider the possibility that the lens may reside in the
Milky Way, in the FRB host galaxy, or in an intervening galaxy. In
Sect.\,2, we briefly introduce the theory and formulae of plasma
lensing. Two models of lensing are shown in Sect.\,3 and Sect.\,4. We
discuss the possible bias plasma lensing introduces to the estimation
of dispersion relation and finally summarize our results.
In this paper we adopt the standard $\Lambda$CDM cosmology with parameters based on the results from the $Planck$ data \citep{2018arXiv180706209P}: $\Omega_{\Lambda}=0.6791$, $\Omega_m=0.3209$, and Hubble constant $H_0=100h$\,km\,s$^{-1}$\,Mpc$^{-1}$ and $h=0.6686$.

\section{Basic formulae of plasma lensing}

The description of gravitational lensing used in this work follows
from \citet{SEF,narayan}. We make the usual thin lens approximation,
which means that we assume weak deflection, and the scattering occurs
only on the lens plane.
We consider a source at angular position $\beta$ with respect to the
line-of-sight, and the corresponding image is formed at the angular position $\theta$. The angular diameter distance from the observer to source, deflecting lens and the difference between are given by $D_s$, $D_d$ and $D_{ds}$, respectively. The lens equation can be written as
\be
\beta = \theta - \alpha = \theta - \nabla_\theta \psi(\theta),
\label{eq:lenseq}
\ee
where $\alpha$ is the deflection angle, $\psi$ is the effective lens
potential and $\nabla_\theta$ is the gradient on the image plane. Lens
models based on both analytical and numerical approaches have also
been explored \citep[e.g.][]{ESE4,FRBplasma1,ErRogers18}. When an
electromagnetic wave propagates through a plasma lens, it will be
delayed due to two separate effects. First, a signal is delayed due to
the increasing path length of propagating along a trajectory that has
been deflected by a plasma lens. This is the geometric component of
the time-delay. Second, an electromagnetic wave propagating through
plasma is also delayed due to dispersion, the frequency-dependent
change of velocity of the signal. This is the effective ``potential''
delay, analogous to the Shapiro delay in gravitational lensing. Unlike
gravitational lensing, both terms of the delay due to plasma are
frequency dependent. In particular the geometric effect, which is
affected by the distribution of the plasma, shows stronger dependence
on the frequency.

We consider the geometric effect due to the light ray being deflected
by a clump of plasma.\footnote{In this work, we only consider the
  geometric optics limit. For the extreme low frequency signal, wave effects
  need to be taken into account.} Then the delay time contributed by
the geometric effect is given by
\citep[e.g.][]{1986ApJ...310..568B,FRBplasma1}.
\be
T_{ge} \simeq {(1+z_d) \over c } {D_d D_s \over D_{ds}} {(\beta-\theta)^2 \over 2}
\elabel{td-geo}
\ee

Next, we consider the contribution from electromagnetic wave
dispersion in plasma. The refractive index of cold plasma for a radio
wave with angular frequency $\omega=2\pi \nu$ is given by
\be
n_{pl}^2\equiv 1- {\omega_p^2\over \omega^2 },
\label{coldPlasma}
\ee
where
\be
\omega_p \equiv \sqrt{{4\pi e^2 n_e \over m_e}}
\ee
is the plasma frequency, $e$ is the electron charge, $m_e$ is the mass of the electron and $n_e$ is the number density of electrons in the plasma.
%
Thus, the delay time due to wave dispersion is given by 
\be
T_{pl} = \int \frac{1}{c}\rund{{1\over  n_{pl}}-1 } \d l
\ee
In general, the plasma frequency is much smaller than the
observational frequency $\omega_p \ll \omega$. The propagation delay time
can be approximated as
\be
T_{pl} \simeq  \int_0^r \dfrac{\omega_p^2 }{2 c\omega^2}\d l
= {2\pi c r_e \over \omega^2} {\rm DM}(\theta)
\elabel{td-pot}
\ee
where $r_e$ is the classical electron radius, and ${\rm DM}(\theta)$
is known as the dispersion measure:
\be
{\rm DM}(\theta)\equiv \int_0^{D_s} n_e(\theta,l) \d l.
\ee
We work with the Born approximation for weak deflection angles,
relevant for both gravitational and plasma lensing in the geometric
optics limit. Thus, the deflection angles are small and the integrals
can be done along unperturbed rays. For great distances, the
dispersion measure is approximated by the projected electron density
along the line of sight, ${\rm DM}(\theta)\approx N_e(\theta)$. In
reality, the dispersive delay is caused by frequency-dependent
refraction of the wave through the inhomogeneous plasma. This causes
radio signals at different frequencies to have different paths and
thus experience different projected density of electrons (DM). While
$N_e$ is the projected density along a straight line of sight. Only
when the electron distribution is uniform, i.e. no deflection
of radio signals occurs, are the two quantities exactly equal. Both
notations will be used interchangeably in this work.

Combining Eq.\,\ref{eq:td-geo} and Eq.\,\ref{eq:td-pot}, the total time delay can be written as the sum of two terms
\be
T(\theta,\beta)= {(1+z_d) \over c}{D_d D_s \over D_{ds}} {(\beta-\theta)^2 \over 2}
+ {2\pi c r_e \over \omega^2 } N_e(\theta).
\elabel{timedelay}
\ee
We define the ``effective plasma lens potential'' by
\be
\psi(\theta) \equiv
\dfrac{1}{(1+z_d)}\dfrac{D_{ds}}{D_d D_s} {\lambda^2 \over 2\pi} r_e N_e(\theta).
\label{eq:psi}
\ee
in a similar fashion to gravitational lensing.
%
%
We will use the geometric term and the dispersive term for the two
contributions to the time delay in this work. 
The geometric delay is proportional to $\alpha^2\propto \lambda^4$ and
is more sensitive to the wavelength than the dispersive delay,
i.e. $\psi \propto \lambda^2$.

For FRBs, the observed contribution to the DM from the lens is summarized in \citep[e.g.][]{yan16}
\be
{\rm DM}_{\rm obs}={\rm DM}_{\rm MW} + {\rm DM}_{\rm IGM} + {\rm DM}_{\rm HG},
\ee
where ${\rm DM}_{\rm MW}$, ${\rm DM}_{\rm IGM}$ and ${\rm DM}_{\rm
  HG}$ denote the contributions from the Milky Way, intervening galaxy
and the host galaxy of the FRB, respectively. The plasma lensing
effects from individual contributions can vary significantly due to
the distance, especially the geometric delay. We will discuss such
effects in the following sections for different models of electron
density and lens distances. In this work, we assume the source is at the
redshift of one repeated FRB \citep{repeatingFRB}, which is
$z_s=0.19273$ ($\sim690.053$ Mpc), and compare the cases of the plasma
lens at different distances: Milky Way ($z=5\times10^{-7}\sim2.24$
kpc), intervening galaxy ($z=0.05\sim210$ Mpc) and FRB host
galaxy ($z=0.192729\sim690.05$ Mpc). Due to the difficulty of both
theory and observation of the ISM on the spatial scales necessary for
ESEs, there are no analytical or empirical expressions for the
detailed density structure of the plasma.
In this work we adopt two analytical forms for the spherically
symmetric electron distribution within a plasma lens which are widely
used in the literature. The exponential models are a family of lenses
that include the most well-known model, the Gaussian lens
\citep[e.g.][]{cleggFey,ESE3,FRBplasma1}, as well as the family of
power-law models \citep{ErRogers18}. These lens families are useful
because they can be used as building blocks to construct more
complicated density distributions.

\section{Exponential model}
\label{sec:Exp}
In this work, we restrict our study to axisymmetric models for the
electron distribution in order to simplify the mathematics and provide
clear, easy to interpret results. Additionally, we only adopt a single
lens along the line of sight.  Exponential lenses are a natural group
of models to consider. The Gaussian lens introduced by
\citet{cleggFey} to describe observations of the extragalactic sources
0954+654 and 1741-038, is a special case of the exponential model
($h=2$). We follow the description of exponential models in
\citet{ErRogers18} and the Gaussian lens in \citet{cleggFey}. We adopt
a form for the electron column density in the lens plane,
\be
N_e(\theta)=N_0\,{\rm exp}\rund{-{\theta^h\over h\sigma^h}}\quad\quad(\theta>0),
\ee
with $N_0$ the maximum electron column density within the lens and
$\sigma$ the width of the lens for $h>0$
\citep{newESE2017}. The projected electron density gives
the potential
\be
\psi(\theta)= \theta_0^2 \exp\left( -\frac{\theta^h}{h\sigma^h} \right)
\label{exp-pot}
\ee
and deflection angle
\be
\alpha(\theta)=-\theta_\text{0}^2
\frac{\theta^{(h-1)}}{\sigma^h}\exp\left( -\frac{\theta^h}{h \sigma^h} \right)
\label{deflExp}
\ee
with the characteristic angular scale
\bea
\theta_0 &= \lambda \left(\frac{D_\text{ds}}{D_\text{s} D_\text{d}}
\frac{1}{2\pi (1+z_d)} r_\text{e} N_\text{0} \right)^\frac{1}{2} \nonumber\\
&= {c \over \nu} \left(\frac{D_\text{ds}}{D_\text{s} D_\text{d}}
\frac{1}{2\pi(1+z_d)} r_\text{e} {\rm DM(0)} \right)^\frac{1}{2},
\elabel{exp-t0}
\eea
where $\lambda$ is the observing wavelength and $\nu$ is the
frequency.  The wavelength of a photon $\lambda$ can vary in the
gravitational field via the gravitational redshift effect, which
introduces an additional complication to the deflection angle. Since
we only focus on lensing from plasma, the gravitational deflection
generated by the ISM will be neglected. The ratio of geometric to
dispersive delay $\eta$ can be given analytically
\be
\eta = {1\over 2}{\alpha^2\over \psi}
=\theta_0^2 {\theta^{2h-2} \over 2\sigma^{2h}}
\exp\rund{-{\theta^h\over h \sigma^h}}.
\ee
In the case of $h\neq1$, the ratio reaches a maximum at
$\theta=(2h-2)^{1/h}\sigma$. The maximum ratio is
\bea
\eta &= \dfrac{(2h-2)^{2h-2\over h}\theta_0^2}{2 \sigma^2}{\rm exp}\rund{-{2h-2\over h}}\nonumber\\
&\propto \dfrac{\lambda^2 }{(1+z_d)} \dfrac{D_{ds}}{D_d D_s}\dfrac{{\rm DM(0)}}{\sigma^2}.
\label{eq:ratio-exp}
\eea
Besides the wavelength, distance, DM and $h$, the ratio is inversely proportional to the width of the lens $\sigma^2$. Since the smaller $\sigma$, the larger the density gradient, the geometric effect becomes stronger.

\begin{figure}
  \centerline{\includegraphics[width=6cm]{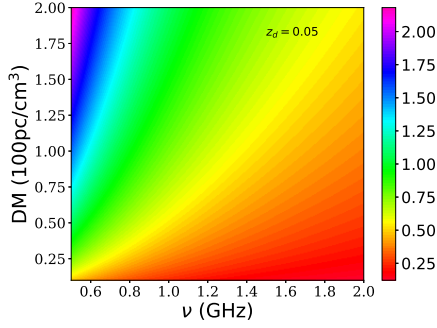}}
  \caption{$\theta_0$ (in unit of milli-arcsec) dependence on
    frequency and DM for the plasma lens of exponential model ($h=2$)
    with lens redshifts $z=0.05$ (intervening galaxy). }
  \label{fig:exp-t0}
\end{figure}

\begin{figure}
  \centerline{\includegraphics[width=4.5cm]{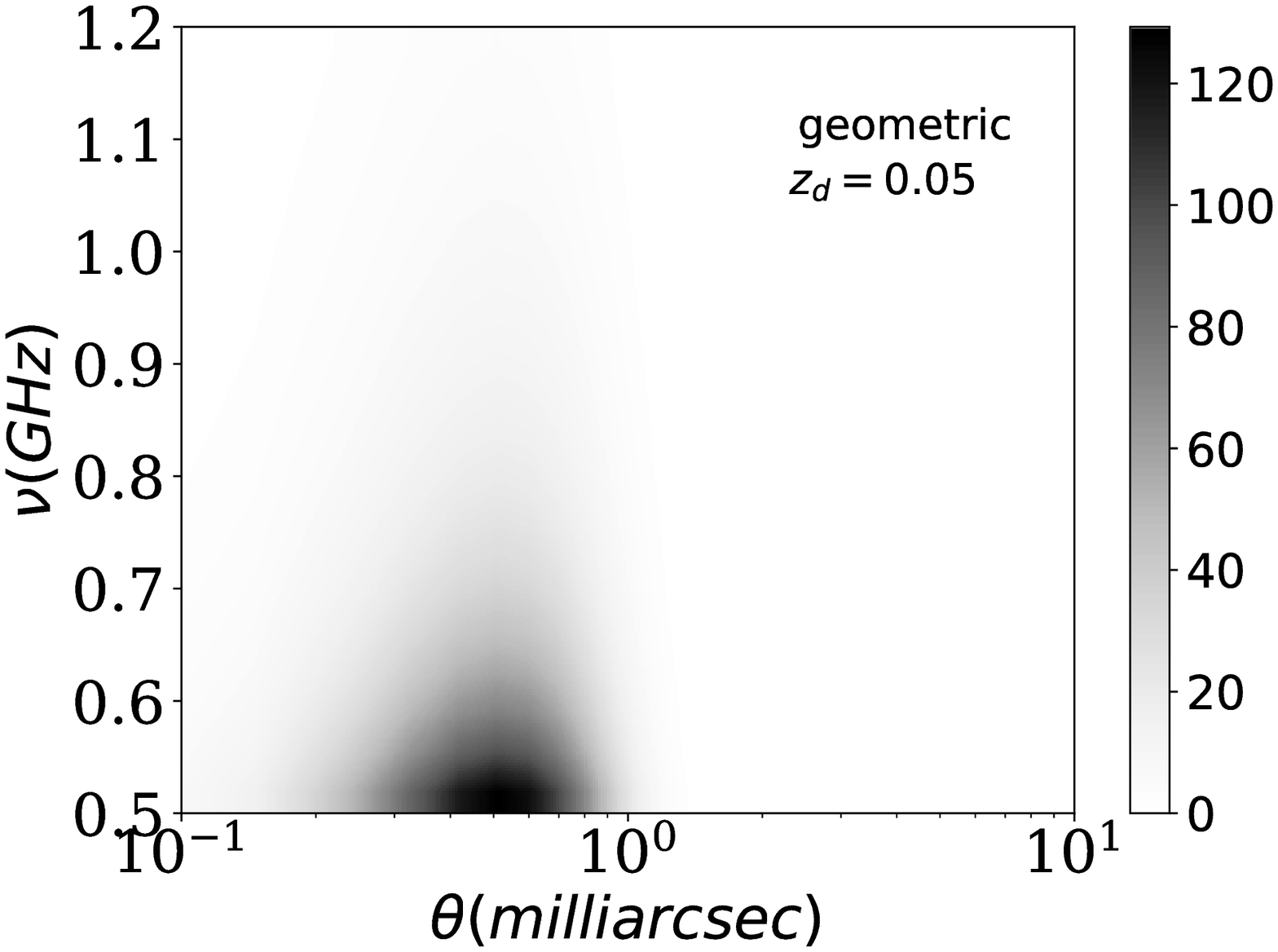}
    \includegraphics[width=4.5cm]{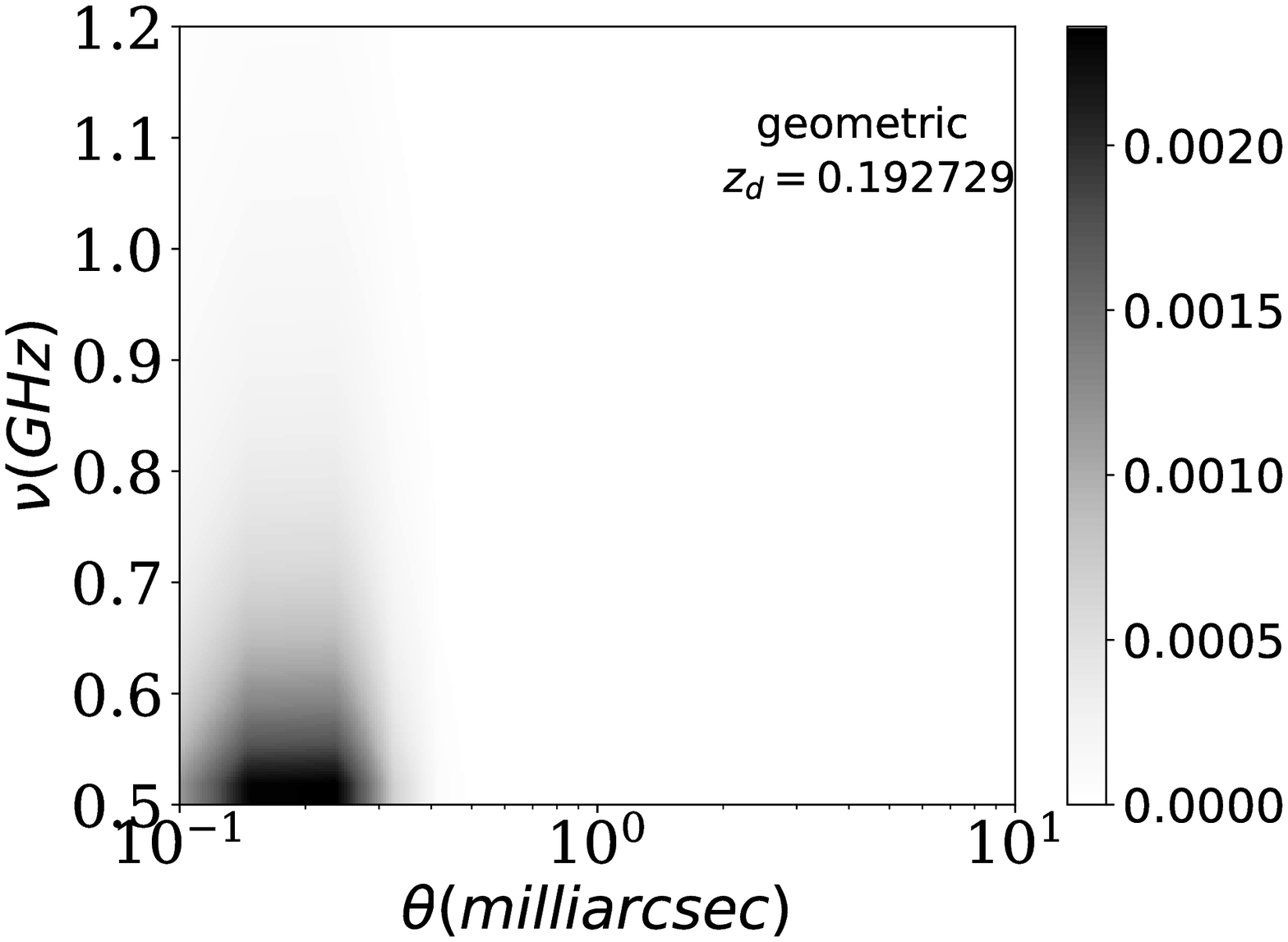}  }
  \centerline{\includegraphics[width=4.5cm]{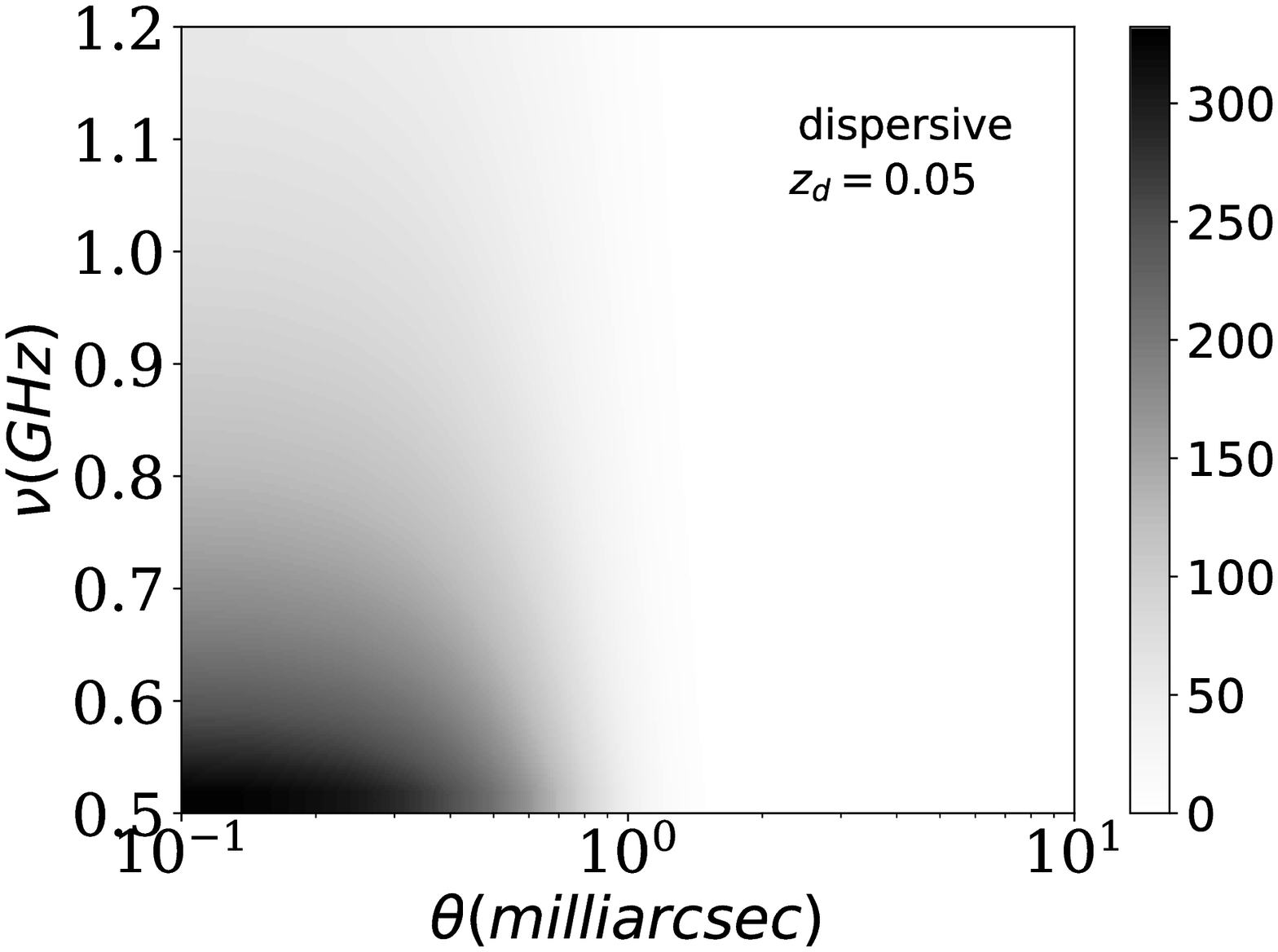}
    \includegraphics[width=4.5cm]{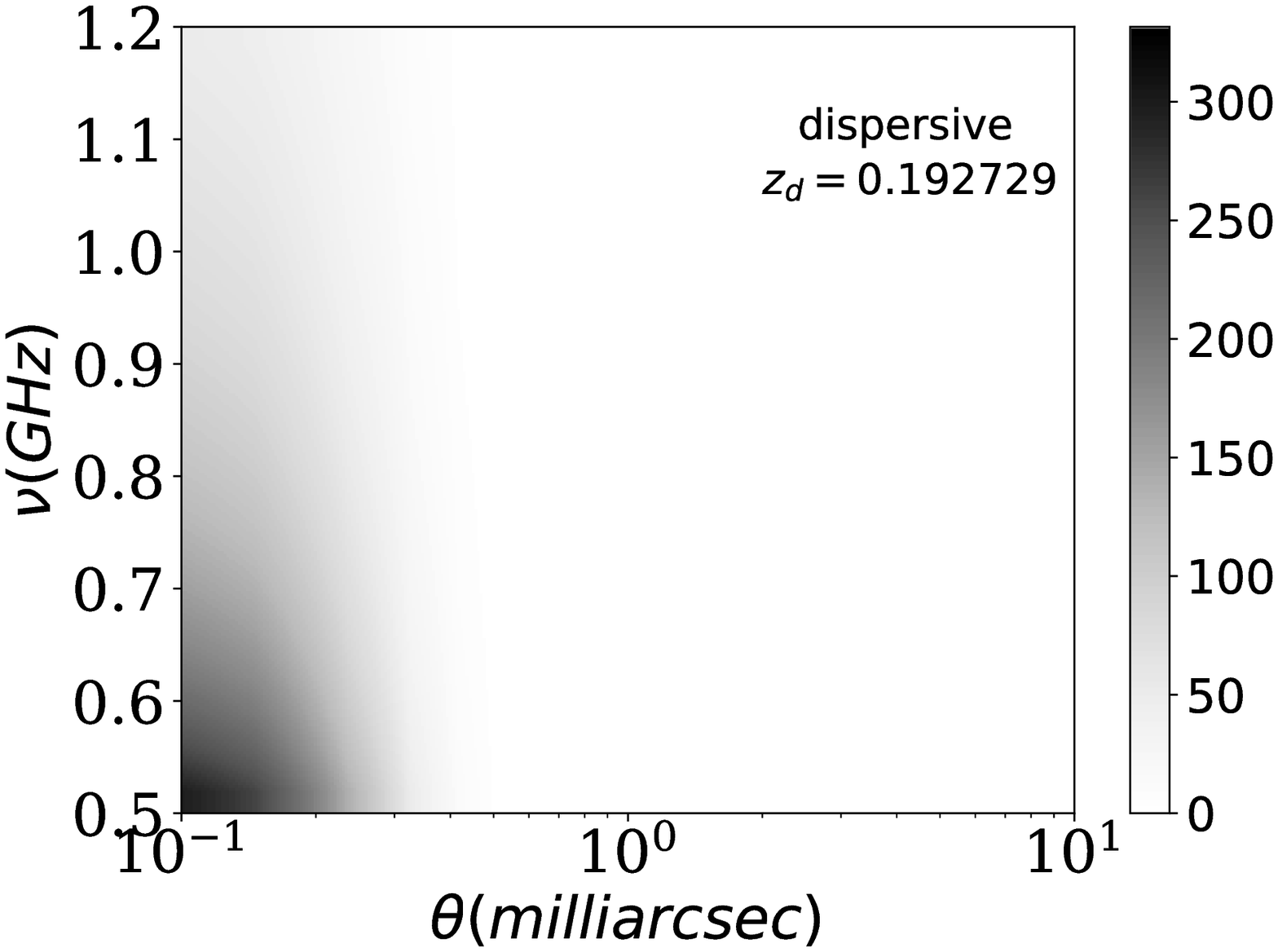}  }
  \centerline{\includegraphics[width=4.5cm]{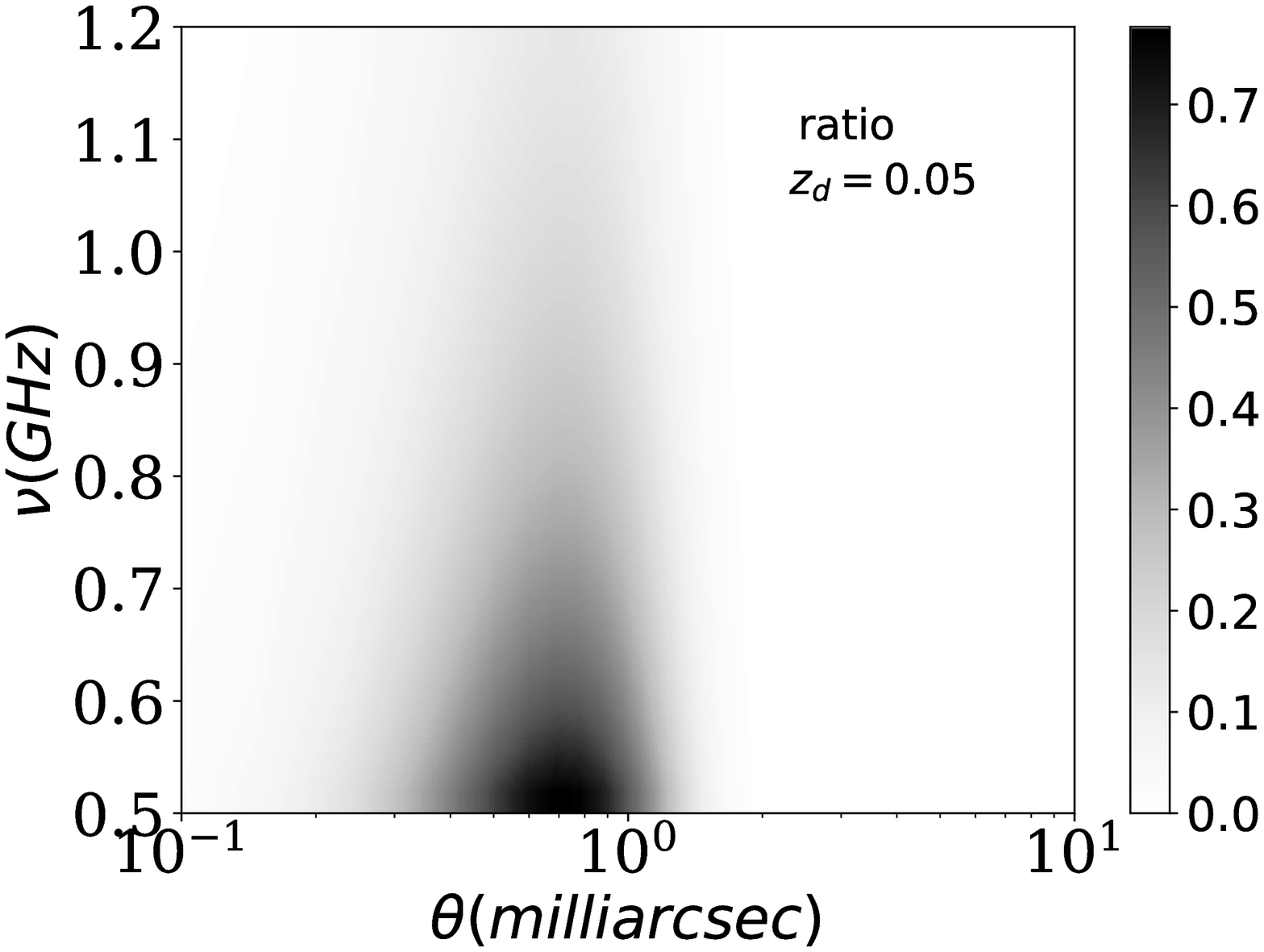}
    \includegraphics[width=4.5cm]{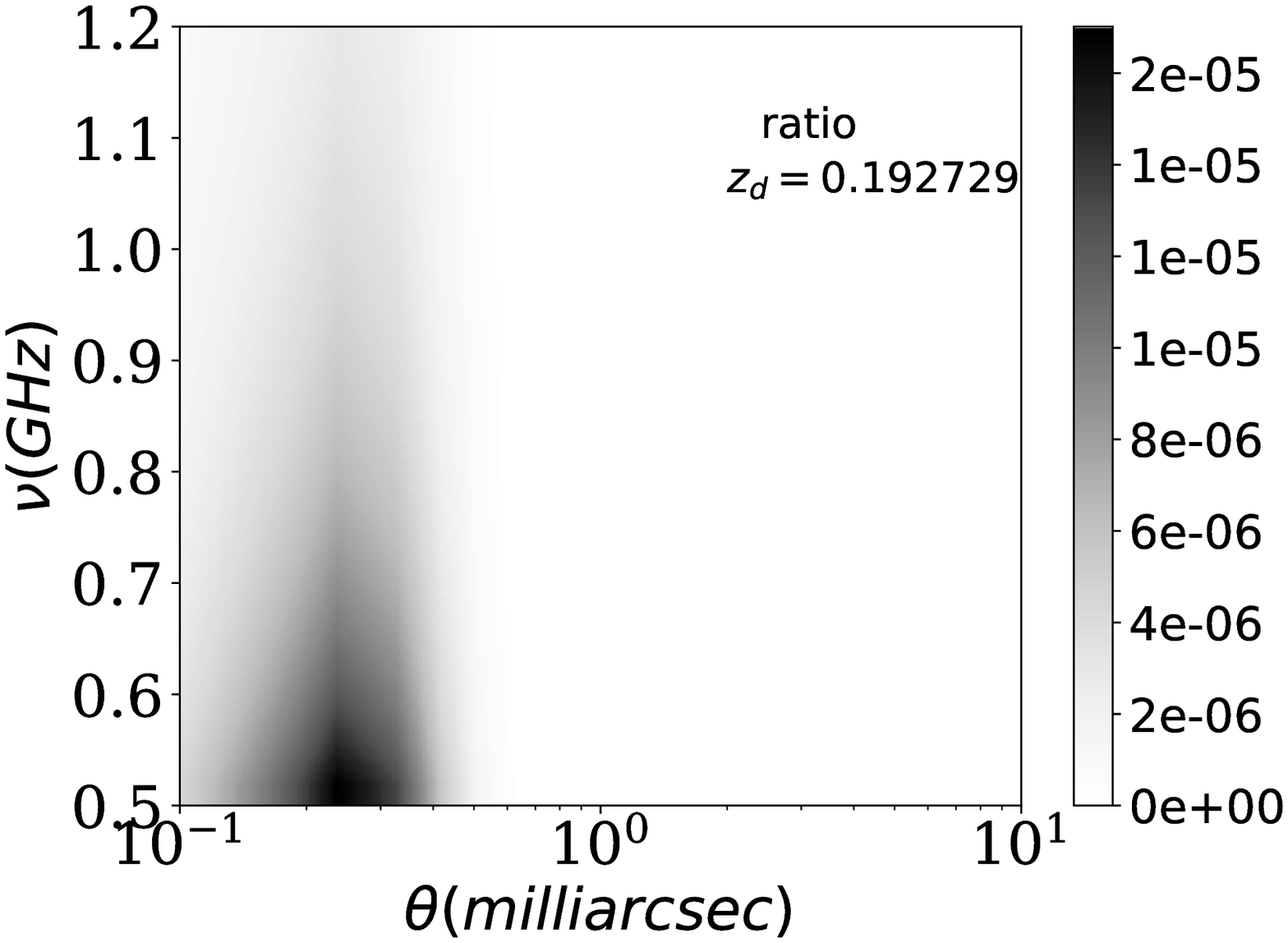}  }
  \caption{The delay time due to the geometric effect 
    (top), dispersive effect (middle), and the ratio between the two
    (bottom) as a function of the image position $\theta$ and
    observational frequency. The time delay is shown in grey scale
    with units of milliseconds. In all the panels, we adopt the same
    exponential plasma lens parameters: $N_0=20$\,pc\,cm$^{-3}$,
    $\sigma=10^5$ AU, $h=2$, but at the different lens redshifts. On
    the left (right) column, the lens redshift is $0.05$
    ($0.192729$). }
  \label{fig:exps1e5}
\end{figure}

The strength of the lens can be characterized by $\theta_0$. The
relationship between $\theta_0$ and $\sigma$ determines the number of
caustic curves in the source plane that separate areas of different
image multiplicity \citep{2019MNRAS.485.5800R}. We note that
$\theta_0$ depends on the frequency of observation, the number density
of electrons as well as the distance of the lens and the source. In
order to clearly see these dependencies, we present $\theta_0$ in
Fig.\,\ref{fig:exp-t0}. In this plot, we give the values of $\theta_0$
for the plasma lens in an intervening galaxy at $z=0.05$. The inverse
dependence of $\theta_0$ on the redshift can easily be seen from
Eq.\,\ref{eq:exp-t0}. Suppose that we choose a large plasma clump to
act as our lens, with $\sigma=10^5$ AU, and a conservative DM range:
$20-200$ pc\,cm$^{-3}$ compared with the observations of FRBs (see the
FRB catalogue)\footnote{In this work, we consider all of the DM to be
  contributed by one plasma structure along the line of sight. In
  reality, all the plasma clumps along the line of sight need to be
  taken into account, which requires a study using multiple lens
  planes. We will leave this more general scenario for future
  work.}. In this case, the average density enhancement within the
lens is a few tens of electrons due to the volume of such a large
lens. A smaller $\sigma$ means the clump is denser, with higher
electron density, but the geometric term in the time delay is stronger
(Eq.\,\ref{eq:ratio-exp}). For lenses at different redshift,
$\theta_0$ will show similar dependence on the observational frequency
and DM, but will have a numerically distinct value with a generally
different order of magnitude. We show the time delay caused by a
plasma lens in Fig.\,\ref{fig:exps1e5}. For lenses at different
redshift, the time delay caused by the dispersive term has the same
order of magnitude, as it is approximately proportional to the
projected electron density, although the angular size of the lensing
region, the lens cross-section, differs significantly due to the
redshift of the lens. On the other hand, the geometric delay shows a
dramatic difference. When the lens is located at the middle point
between us and the source, the geometric effect can contribute a time
delay of similar order to that of dispersive delay, but has different
dependence on the image position ($\theta$), i.e. the delay time
caused by the dispersive term reaches the maximum at the peak of the
projected density, and the delay time caused by the geometric term
reaches the maximum when the gradient of the density peaks. The ratio
of geometric to the dispersive delay is shown in the bottom panel,
demonstrating under which conditions it is safe to neglect the
geometric delay. When the lens is in the Milky Way (the host galaxy),
the geometric term becomes much smaller (see the right panel in
Fig.\,\ref{fig:exps1e5}). If the density gradient is large, the
geometric delay can contribute a significant part as well.

\begin{figure}
  \centerline{\includegraphics[width=4.5cm]{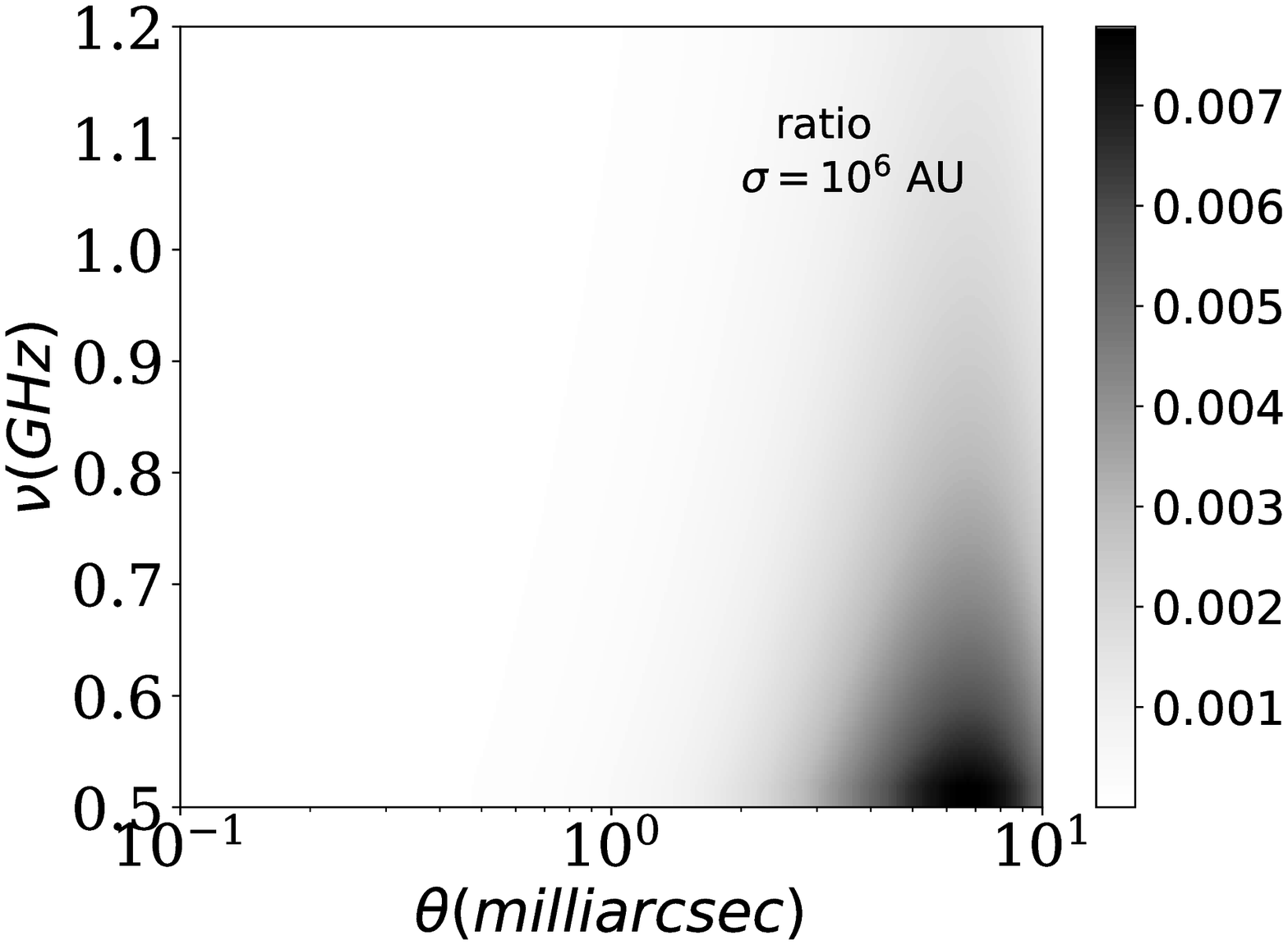}
    \includegraphics[width=4.5cm]{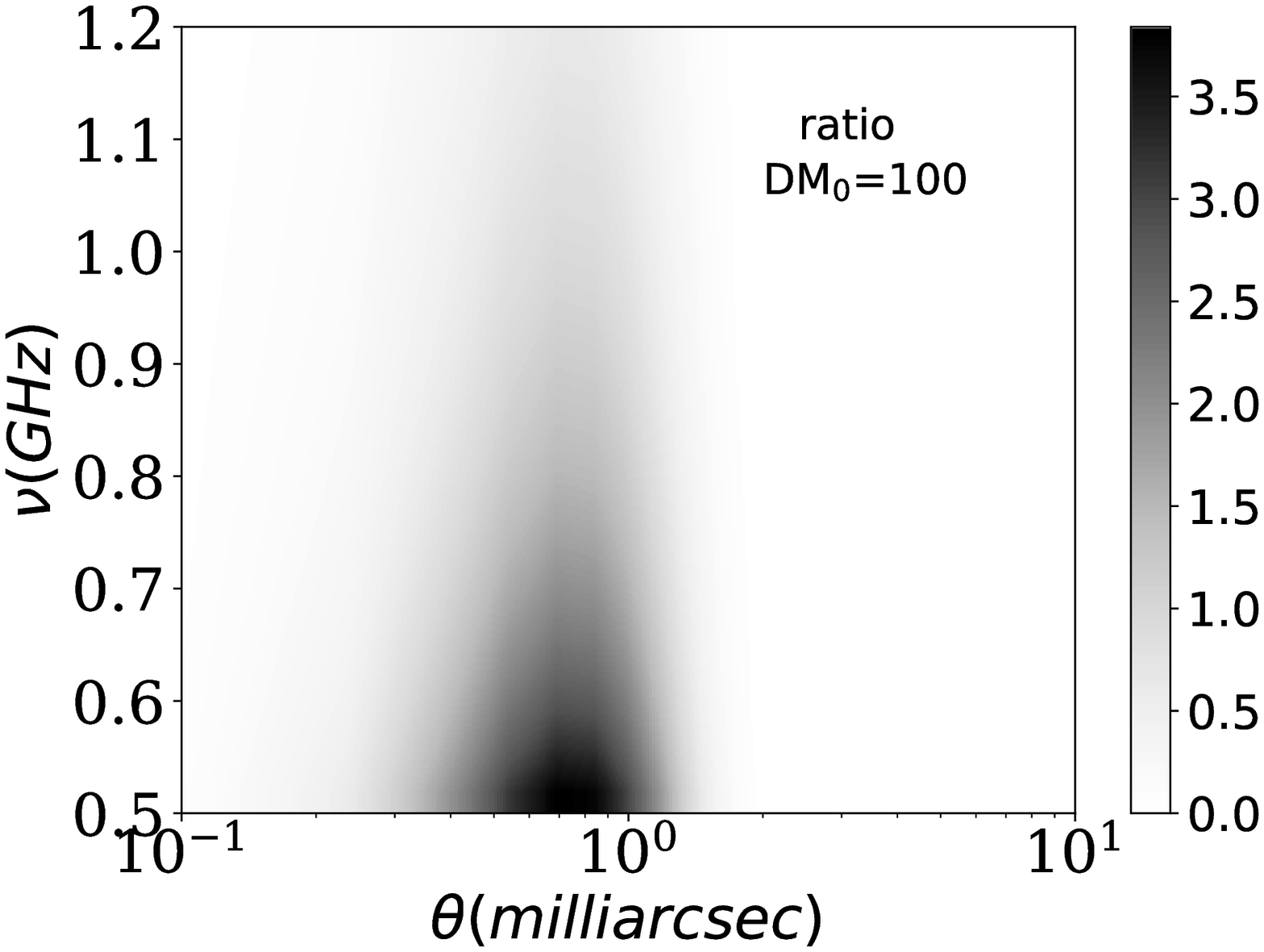}}
  \centerline{\includegraphics[width=4.5cm]{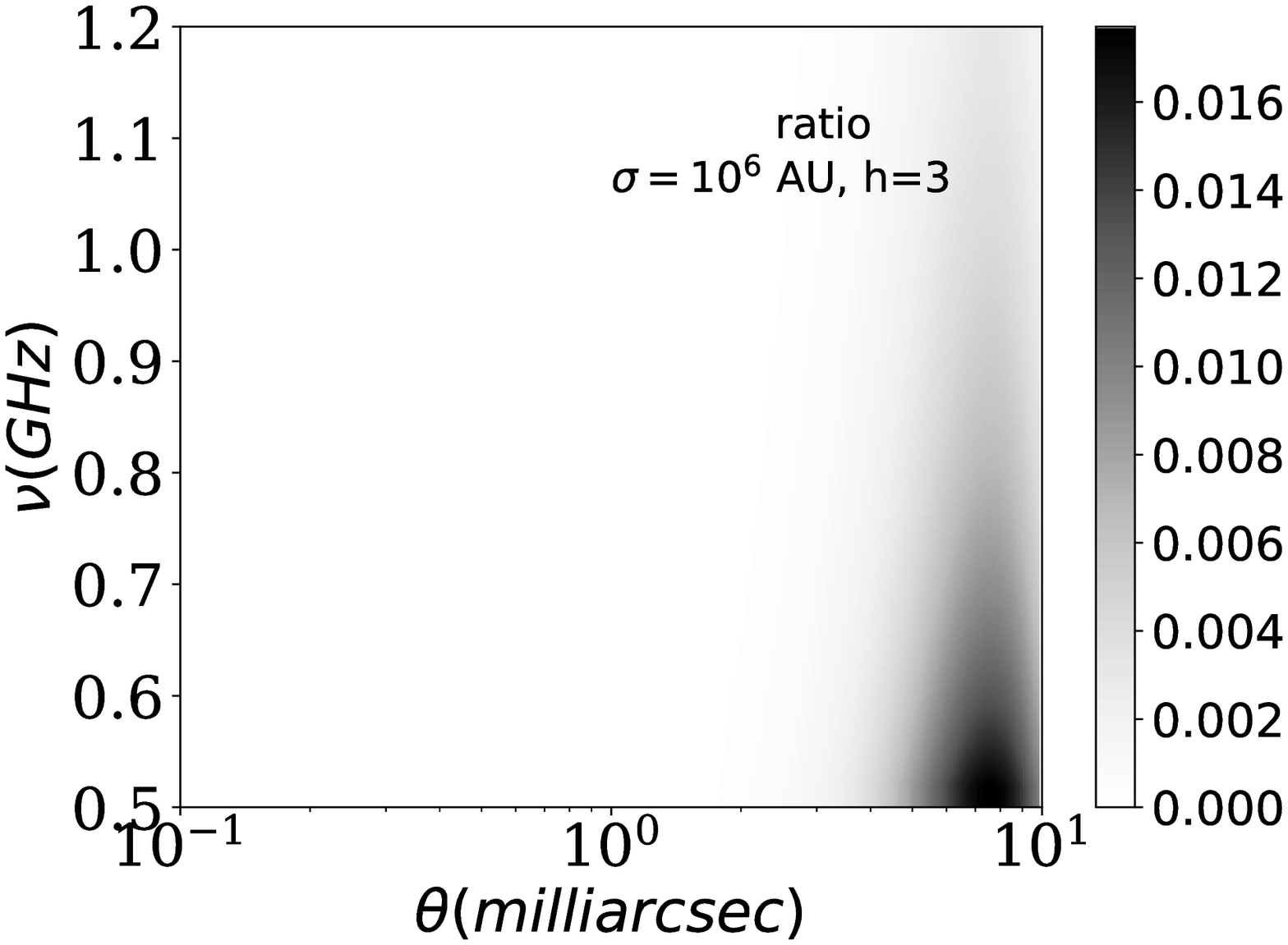}
    \includegraphics[width=4.5cm]{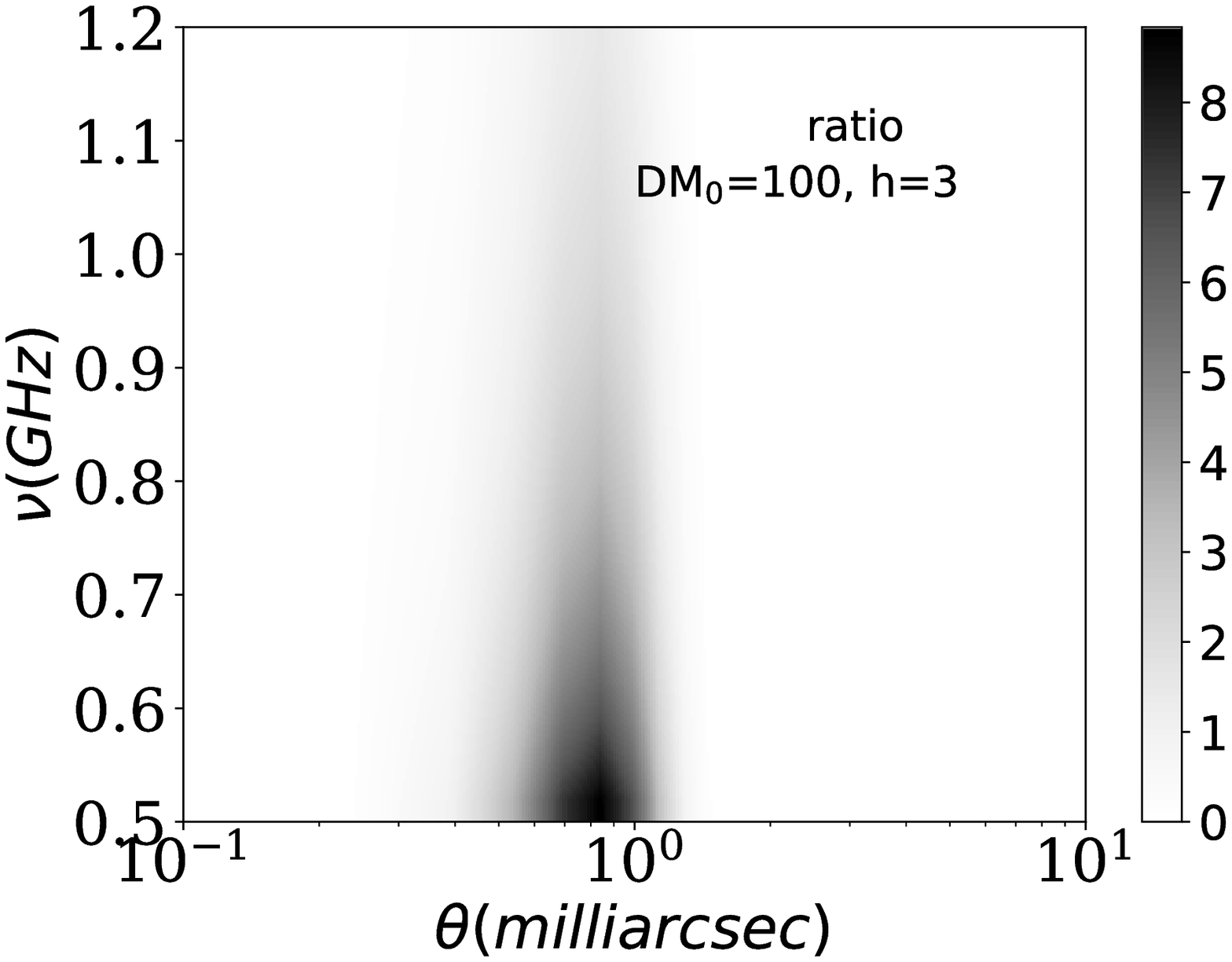}}
  \caption{The ratio of geometric to dispersive delay for a lens at
    $z_d=0.05$. The same lens parameters are adopted as
    Fig.\,\ref{fig:exps1e5} except that given in the corner of each
    panel. }
  \label{fig:exps1e6}
\end{figure}

\begin{figure}
  \centerline{\includegraphics[width=4.5cm]{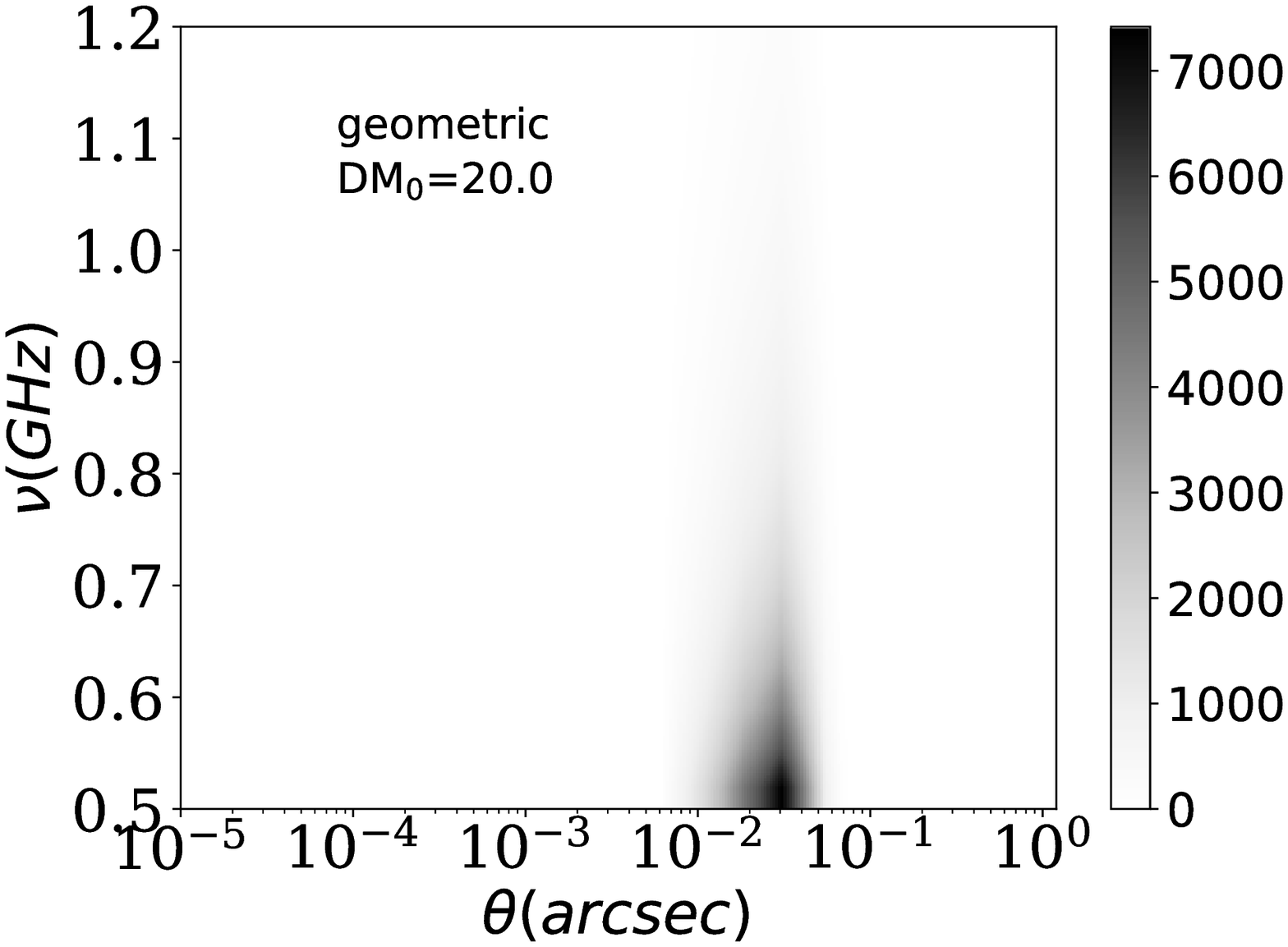}
    \includegraphics[width=4.5cm]{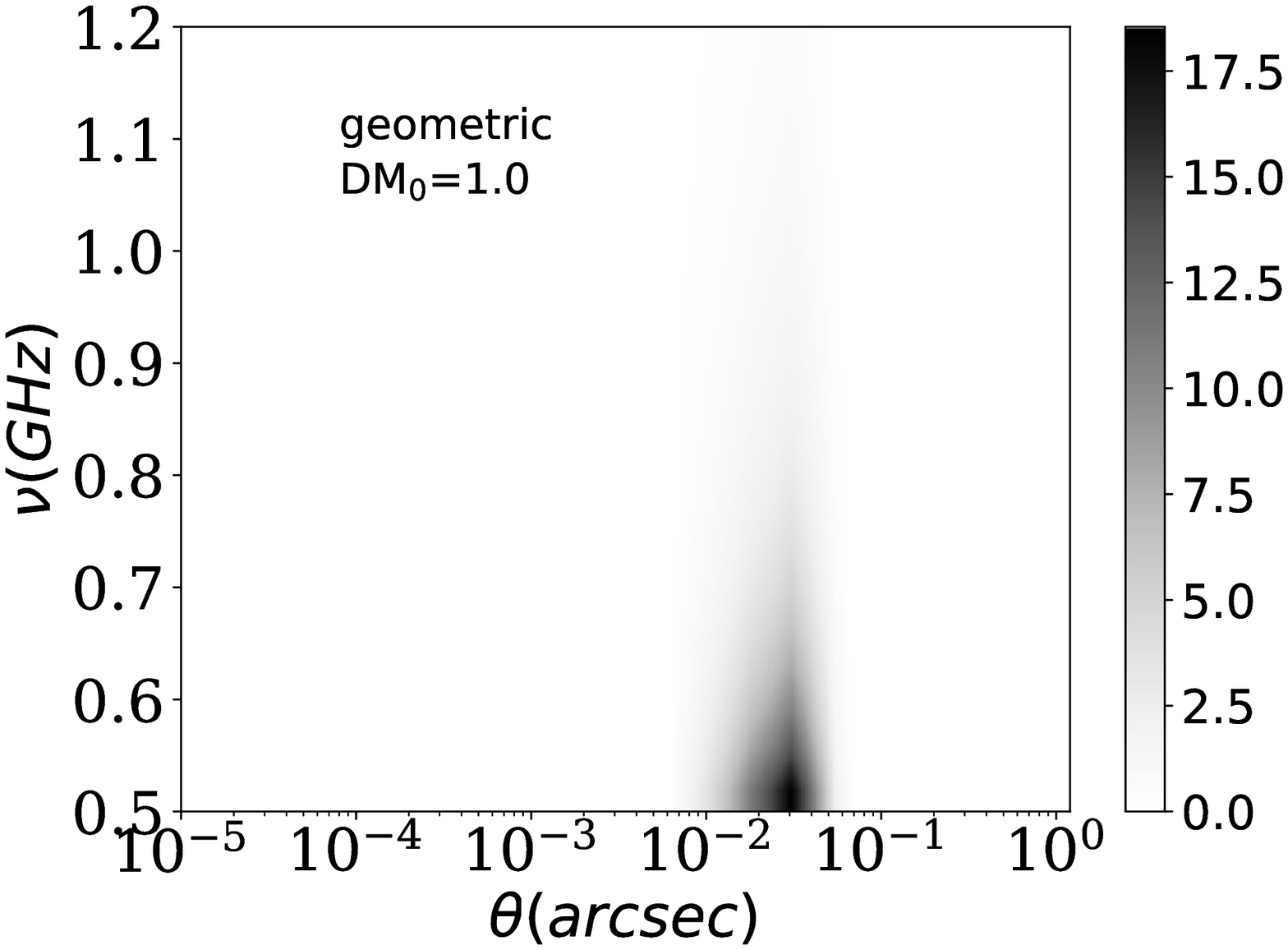}}
  \centerline{\includegraphics[width=4.5cm]{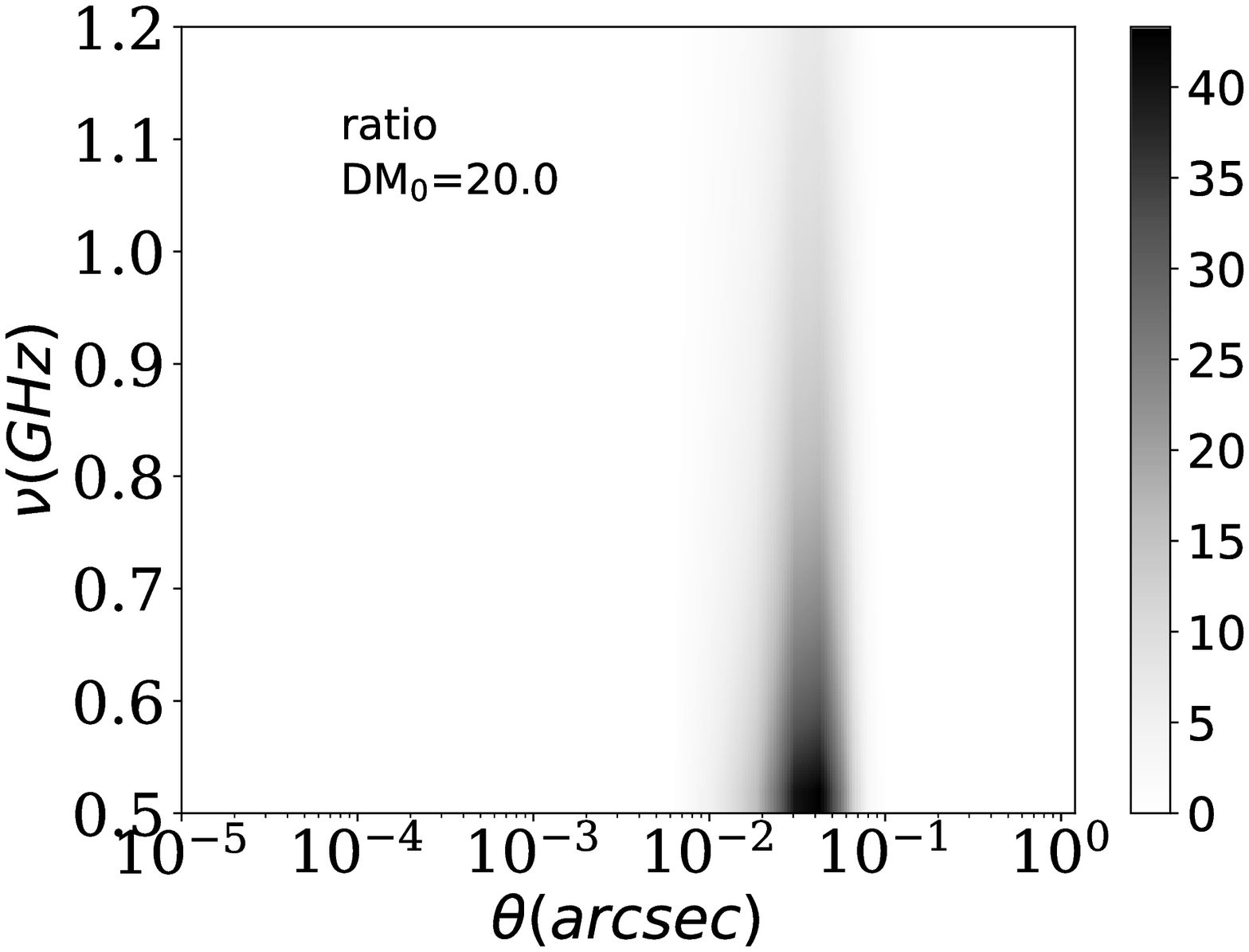}
    \includegraphics[width=4.5cm]{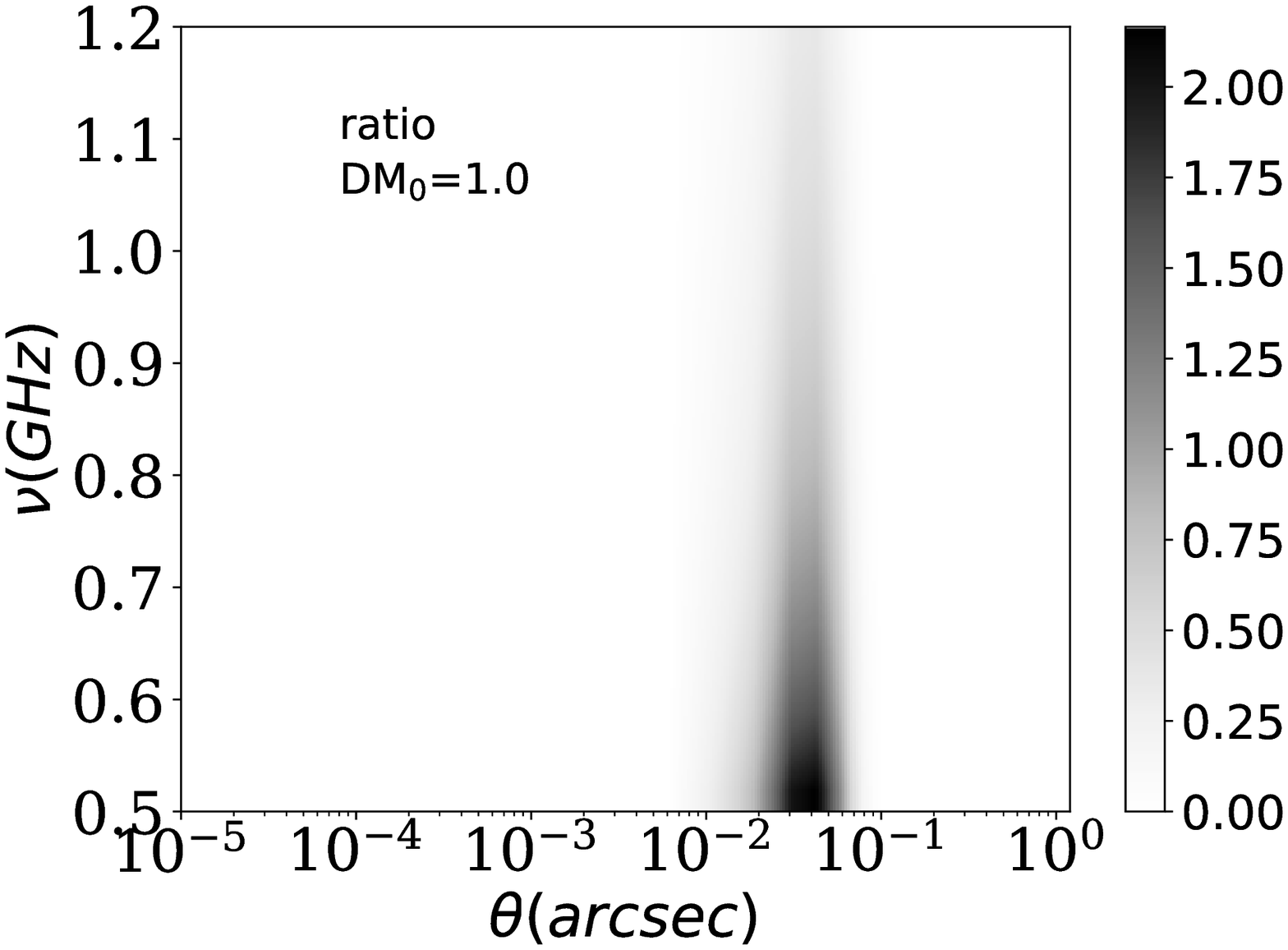}}
  \caption{The same as Fig.\,\ref{fig:exps1e5} for plasma lenses at
    the Milky Way. Different lens parameters ($\sigma=50$ AU and DM is
    given in each panel) are adopted in this figure. }
  \label{fig:exptdlz0}
\end{figure}

The geometric term strongly depends on the model parameters of the
lens, in our case the width $\sigma$ and charge density $N_0$.  From
Table\,\ref{tab:exp-para}, the lens parameters affect the two delay
terms in different ways. Another interesting point is that the
geometric delay is proportional to $\lambda^4$, which is independent
of the lens properties. In Fig.\,\ref{fig:exps1e6}, the ratio between
two delays for lenses at $z=0.05$ are shown. As expected, when the
gradient becomes significantly large, the geometric term can dominate
over the dispersive term. A plasma lens in the Milky Way is slightly
different from the other two cases. Firstly the probability that the
FRB signal propagates through a clump of plasma in the Milky Way is
high. Secondly, the small spatial variations of the plasma clump can
cause substantial lensing effects. It has been suggested that the
scale of the ISM clumps varies from 0.1 AU to a few hundred AU
\citep{2018ARA&A..56..489S}. We thus choose a small scale lens with
$\sigma=50$ AU, and show two separate plots for lenses within our
Galaxy in Fig.\,\ref{fig:exptdlz0}. The units for the horizontal axis
are different from previous figures, since the cross-section of these
lenses are large. Due to the small scale variations in the plasma,
even for a small DM$_0=1$, the geometric delay can essentially equal
the dispersive delay.


\begin{table}
  \caption{The {\it approximate} dependence of two time delay terms on the parameters of the exponential lens model.}
  \begin{tabular}{c|c|c}
    parameter  &geometric term  &dispersive term\\
    \hline
               & $\propto \alpha^2$    &$\propto \psi$\\
    $N_0$      & $\propto N_0^2$       & $\propto N_0$\\
    $\lambda$  & $\propto \lambda^4$   & $\propto \lambda^2$\\
    $\sigma$   & $\propto 1/\sigma^{2h}$   & $\propto {\rm exp}\rund{-1/(h\sigma^h)}$\\
    \hline
  \end{tabular}
  \label{tab:exp-para}
\end{table}
%

\section{Power-law model}
\label{sec:spl}
The power-law (PL) model serves as a useful example for the density
profiles in gravitational lensing \citep{2001astro.ph..2341K} and plasma lensing \citep{ErRogers18}. This is not only due to its well studied analytical behaviour, but also because combinations of PL profiles can be built to mimic other more complex profiles and give rise to interesting optical properties. The three dimensional electron density is given by
\be
n_e(r)=n_0 {R_0^h \over r^h},
\ee
where $n_0$ is the density at radius $r=R_0$. The correponding DM$_0$
can also be given for the projected density at $r=R_0$. However,
usually $R_0/D_d$ is much larger than $\theta_0$ so DM$_0$ does not
describe the density where we are interested. In order to avoid the
singularity at the center, we include a finite core with angular core
radius $\theta_C$. The plasma lens potential of the softened power-law
(SPL) can thus be written as
\be
\psi(\theta) = \dfrac{\theta_{0}^{h+1}}{(h-1)} \dfrac{1}{\left(\theta^2 + \theta_C^2\right)^\frac{h-1}{2}}, h \neq 1
\label{spl_potential}
\ee
with the characteristic angular scale \citep{BKT09,BKT15}
\begin{equation}
\theta_0 = \left( \lambda^2 \frac{D_\text{ds} }{(1+z_d)D_\text{s} D_\text{d}^h} \frac{r_\text{e} n_0 R_0^h }{\sqrt{\pi}} \frac{\Gamma\left( \frac{h}{2} + \frac{1}{2} \right)}{\Gamma\left( \frac{h}{2} \right)} \right)^\frac{1}{h+1}.
\label{spl_t0}
\end{equation}
The SPL lens with $h=1$ and the point-like plasma lens model have
different forms of the potential. The SPL potential gives the deflection angle
\begin{equation}
\alpha(\theta) = - \theta_0^{h+1} \frac{\theta}{\left( \theta^2 + \theta_C^2\right)^\frac{h+1}{2}}.
\end{equation}
The core radius $\theta_C$ can cause complicated behavior of the SPL
lens. In this work, we simply choose $\theta_C=0.05R_0/D_d$ unless we otherwise
specify. One can find more detail about the SPL lens in \citet{ErRogers18}.

The ratio of geometric to dispersive delay can be also given analytically
\be
\eta=\dfrac{(h-1) \theta_0^{h+1} \theta^2}{\rund{\theta^2 + \theta_c^2}^{h+3\over 2}}.
\ee
The maximum ratio will be reached at $\theta=\sqrt{2/(h+1)}\theta_C$,
\be
\eta=\dfrac{(h-1) \theta_0^{h+1} (h+1)^{h+1\over 2}}
           {(h+3)^{h+3\over 2} \theta_C^{h+1}}
           \propto \dfrac{\lambda^2 n_0R_0^h}{\theta_C^{h+1}}.
\ee
A small core radius $\theta_C$ will cause two effects. First, the large geometric delay will only appear at small radius, i.e. describes a small cross section near the lens. On the other hand, when the density gradient becomes large the geometric delay will be large as well.

\begin{figure}
  \centerline{\includegraphics[width=6cm]{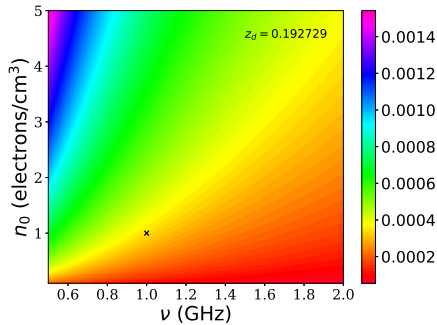}}
  \caption{$\theta_0$ for the PL model with $h=1$, $R_0=10^6$ AU. The
    black cross marks the value given in Table\,\ref{tab:spl-t0}. In
    this lens configuration, $n_0=1$cm$^{-3}$ corresponds to column
    density DM$_0\sim5$pc\,cm$^{-3}$.}
  \label{fig:spl-t0}
\end{figure}

We present the dependence of $\theta_0$ (Eq.\,\ref{spl_t0}) on the
frequency and the number density of electrons in
Fig.\,\ref{fig:spl-t0}. It shows a similar pattern to the exponential
models. The magnitude of $\theta_0$ also varies according to the lens
redshift and power index. We list the value of $\theta_0$ in
Table\,\ref{tab:spl-t0} for a set of selected parameters ($n_0=1$
cm$^{-3}$, $\nu=1$ GHz, $R_0=10^6$ AU).
\begin{table}
  \caption{$\theta_0$ (milli-arcsec) for PL model of lens with $n_0=1\,{\rm cm}^{-3}$, $\nu=1$GHz, $R_0=10^6$ AU.}
  \begin{tabular}{c|c|c|c}
      &$z=5e-7$  &$z=0.05$  &$z=0.192729$\\
    \hline
    $h=1$   &86.5 &0.241  &3.46e-4 \\
    $h=2$   &1741  &0.758  &6.49e-3  \\
    $h=3$   &7412   &1.28  &0.0267  \\
    \hline
  \end{tabular}
  \label{tab:spl-t0}
\end{table}

\begin{figure}
  \centerline{\includegraphics[width=4.5cm]{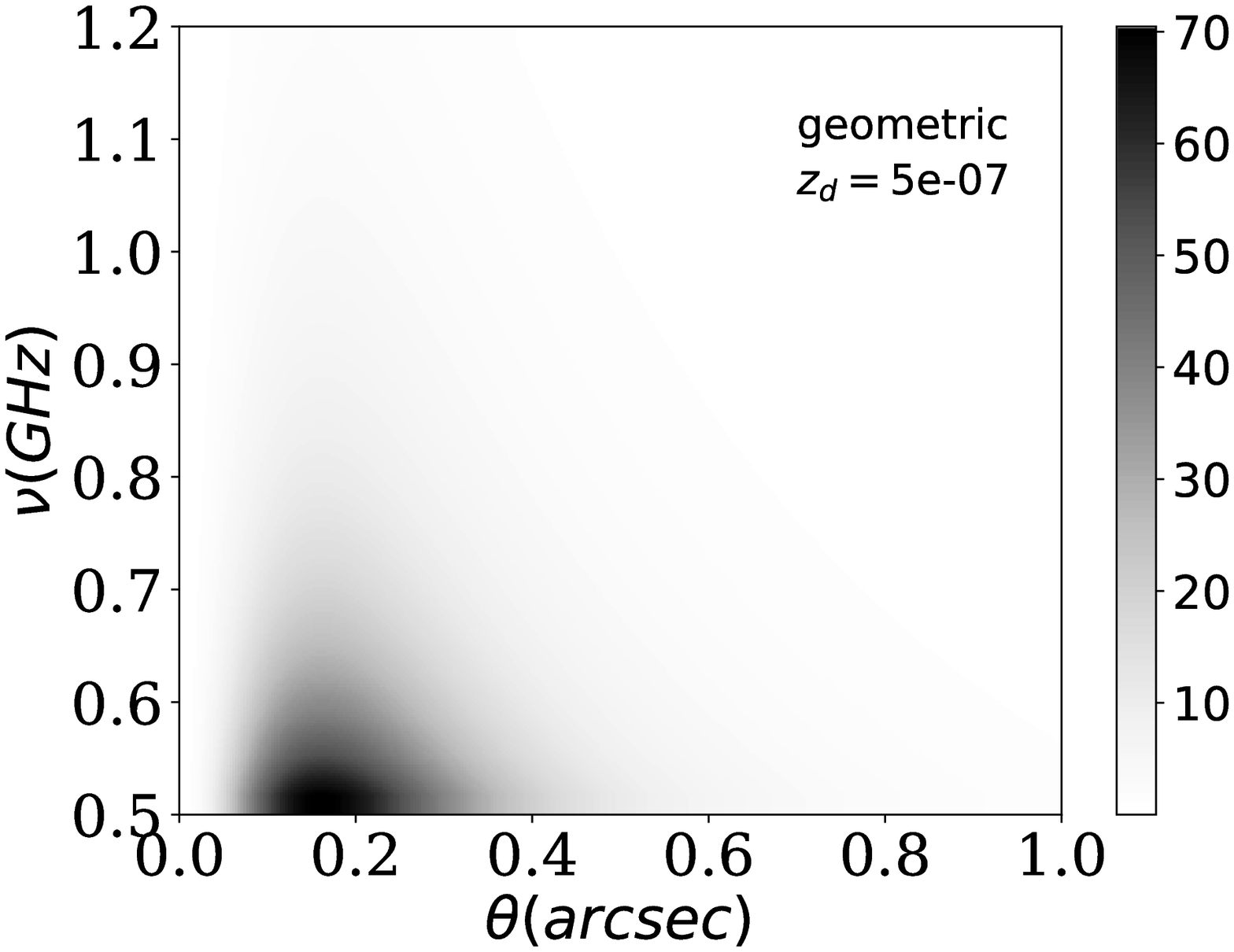}
    \includegraphics[width=4.5cm]{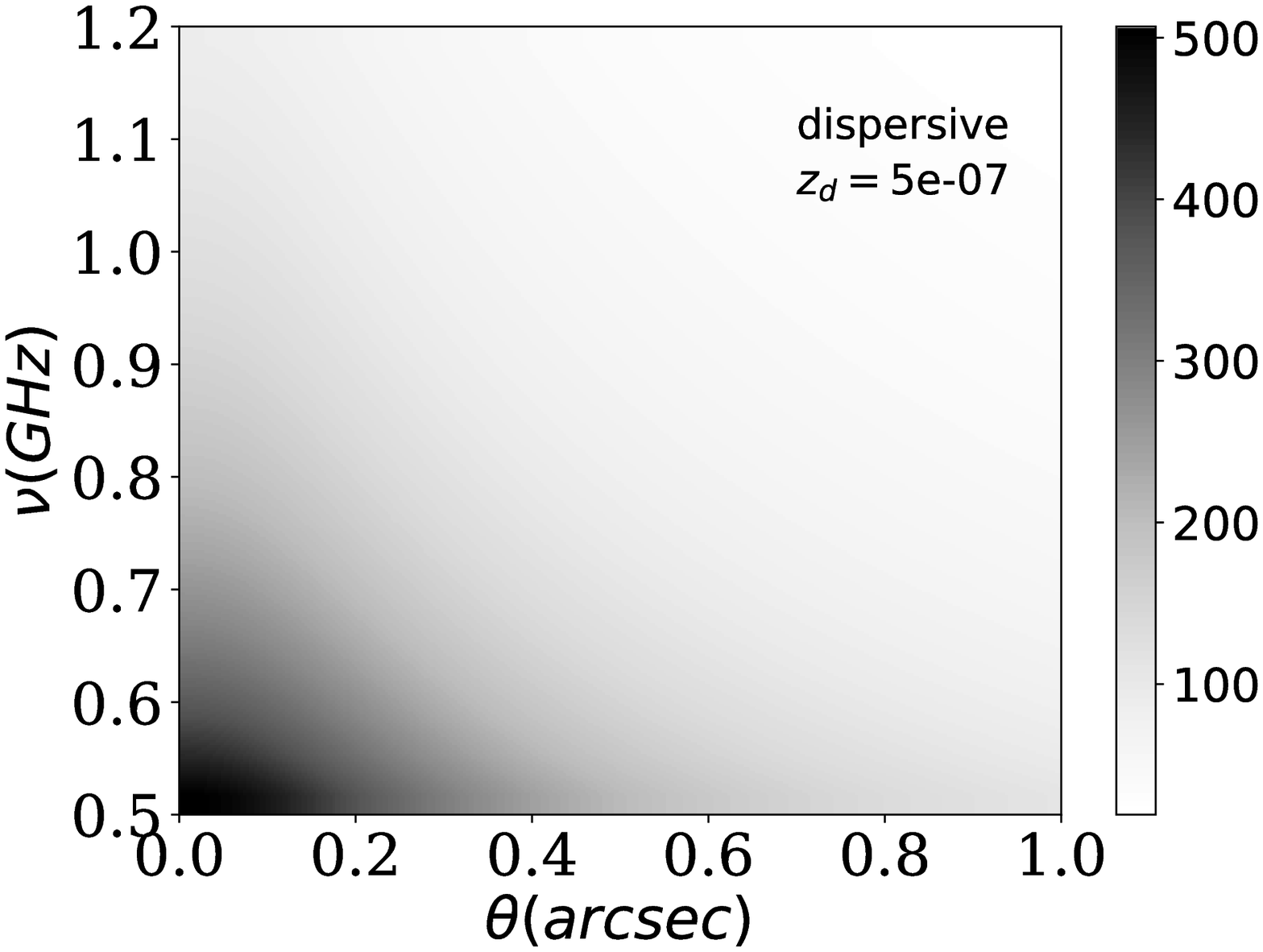}}
  \centerline{\includegraphics[width=4.5cm]{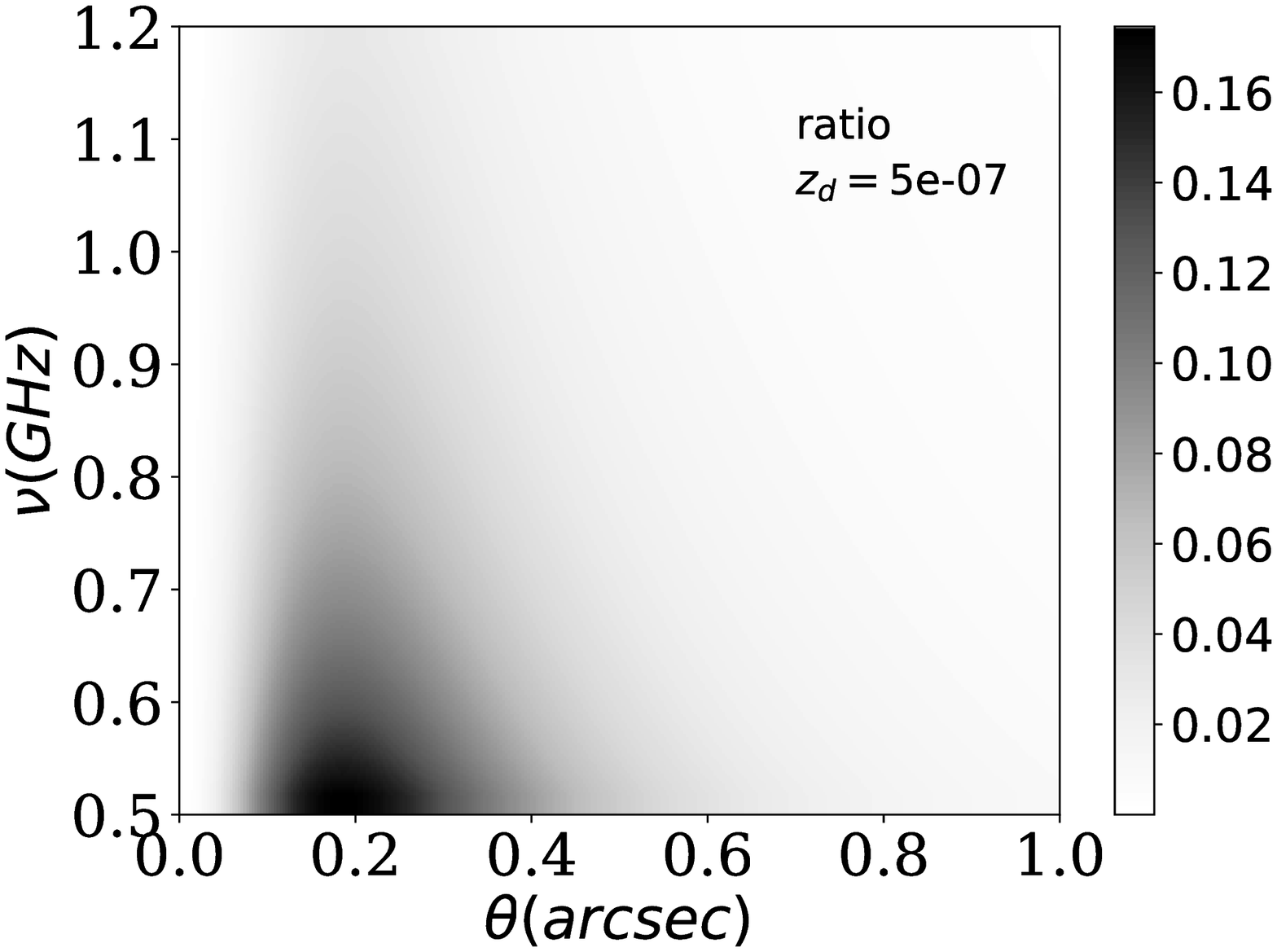}}
  \caption{The time delay due to the plasma lensing geometric effect
    (left), the dispersive effect (right), and their ratio for an SPL
    model.  The lenses are located in Milky Way $z=5e-7$ with lens
    parameters: $h=2$, $n_0=0.1$cm$^{-3}$, $R_0=10^5$ AU, and
    $\theta_C=0.005R_0/D_d$. In this figures for lens in the
      Milky Way, the separation between the lens and source is shown
      in linear scale for better visibility.}
  \label{fig:spl-tdl}
\end{figure}

\begin{figure}
  \centerline{\includegraphics[width=4.5cm]{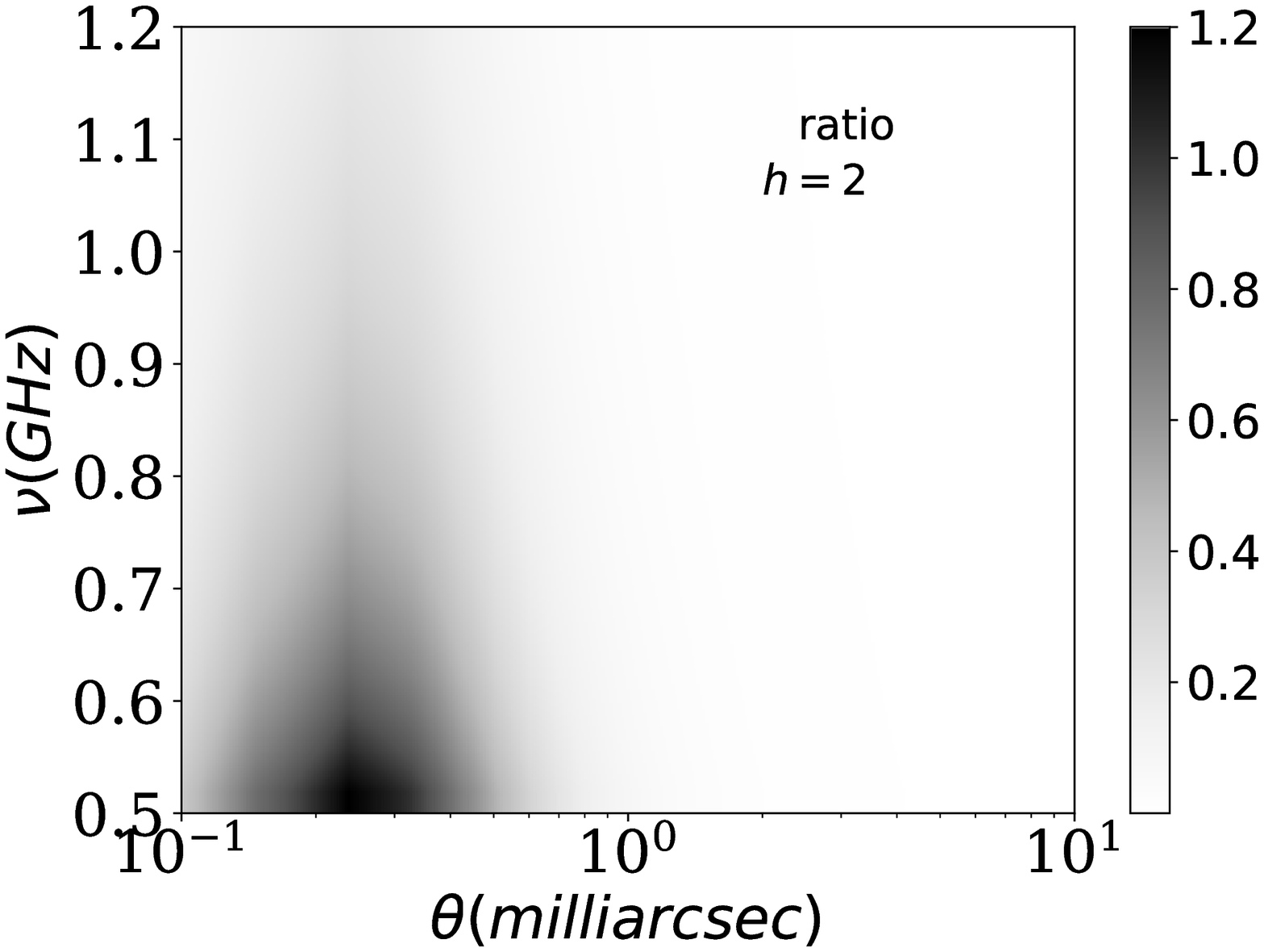}
    \includegraphics[width=4.5cm]{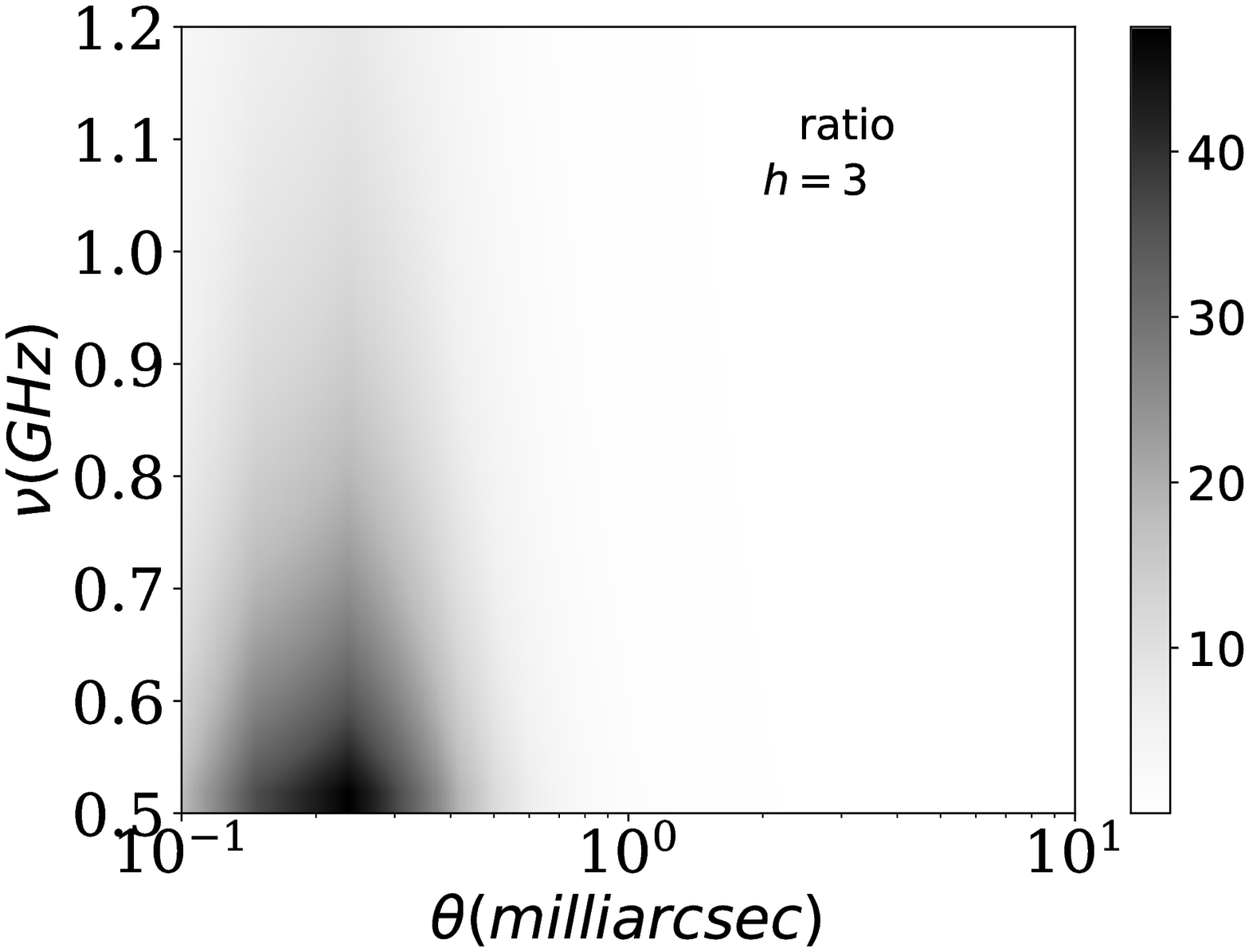}}
  \caption{Same as Fig.\,\ref{fig:spl-tdl} but for lenses at $z=0.05$
    (intervening galaxy). The lens parameters are
    $n_0=0.1$ cm$^{-3}$, $R_0=10^6$ AU, and the power index is given in
    each panel. }
  \label{fig:spl-tdl2}
\end{figure}

We also compare the time delay between the geometric and dispersive
terms for SPL models. In Fig.\,\ref{fig:spl-tdl}, we present a lens in
our Galaxy. The projected density at $R_0$ is
$0.05$\,pc\,cm$^{-3}$. While near the central region of the lens
(within a hundred AU), the density dramatically increases up to a
thousand electrons per cm$^{3}$, and DM reaches
$\sim50$\,pc\,cm$^{-3}$. Such a high electron density will not
increase the overall average density but can generate a large density
gradient and cause strong lensing effects as well as a large geometric
time delay. Especially when the plasma lens is in the intervening
galaxy (Fig.\,\ref{fig:spl-tdl2}), the geometric delay can dominate
over the dispersive delay. The lens parameters play a critical role
for the geometric delay, but one has to be careful that the relation
given in the Table\,\ref{tab:spl-para} is for PL lens models. A large
core radius can totally change the dependence, as it softens the
density gradient, and reduces the lensing effect \citep{ErRogers18}.

We compare the two regions on the lens plane: the first where the delay ratio is greater than $0.01$, and second where the delay ratio is greater than $0.1$. In general these regions depend on the frequency of the observations. What we will present here is the cross-section at $\nu=0.5$\, GHz. For the Gaussian lens (Fig.\,\ref{fig:exptdlz0}), the fraction is about $0.56$ for DM$_0=20$\,pc\,cm$^{-3}$ and $0.7$ for DM$_0=100$\,pc\,cm$^{-3}$. For the SPL lens in Fig.\,\ref{fig:spl-tdl}, the fraction is about $0.14$. If we assume that FRBs are uniformly distributed behind the lens, the fraction can be used to estimate the probability that the geometric effect will contribute a non-negligible delay. It varies significantly with the lensing properties. For the plasma clump with small scale variations, the contribution from the geometric effect is high. 


%
\begin{table}
  \caption{The dependence of two time delay terms on the paramters of the PL lens model.}
  \begin{tabular}{c|c|c}
    parameter  &geometric term  &dispersive term\\
    \hline
    lens para.  &$\propto n_0^2R_0^{2h}$  & $\propto n_0R_0^h$ \\
    $\lambda$   &$\propto \lambda^4$     & $\propto \lambda^2$ \\
    \hline
  \end{tabular}
  \label{tab:spl-para}
\end{table}
%

\section{Bias in estimating the Dispersion Measure}

The frequency-dependent time delay can be used to estimate the DM by
fitting the frequency-time delay curve of compact radio sources
\citep[e.g.][]{petroff2016}. However, it depends on the assumption
that the density gradient can be neglected, i.e., if the density
gradient causes deflection of the background radio signal, the
frequency-time delay relation will diverge from the general one.  Such
effects have been noticed in the study of pulsars
\citep[e.g.][]{2016ApJ...817...16C,2018Natur.557..522M}. As we will
see it is also significant in propagation of FRBs. For the two delay
terms in Eq.\,\ref{eq:timedelay}, the dispersive term approximately
equals the DM. The geometric effect, which is also frequency-dependent
($\propto \lambda^4$), will not only cause time delay of the signal,
but also change the trajectory of the radio signal. Therefore, the DM
that the signal experiences during the propagation will also change
according to the frequency. This is also the reason why the dispersive
term does not amount to the entire DM. In the end, the estimated DM
without plasma lensing is thus biased.

\begin{figure}
  \centerline{\includegraphics[width=8cm]{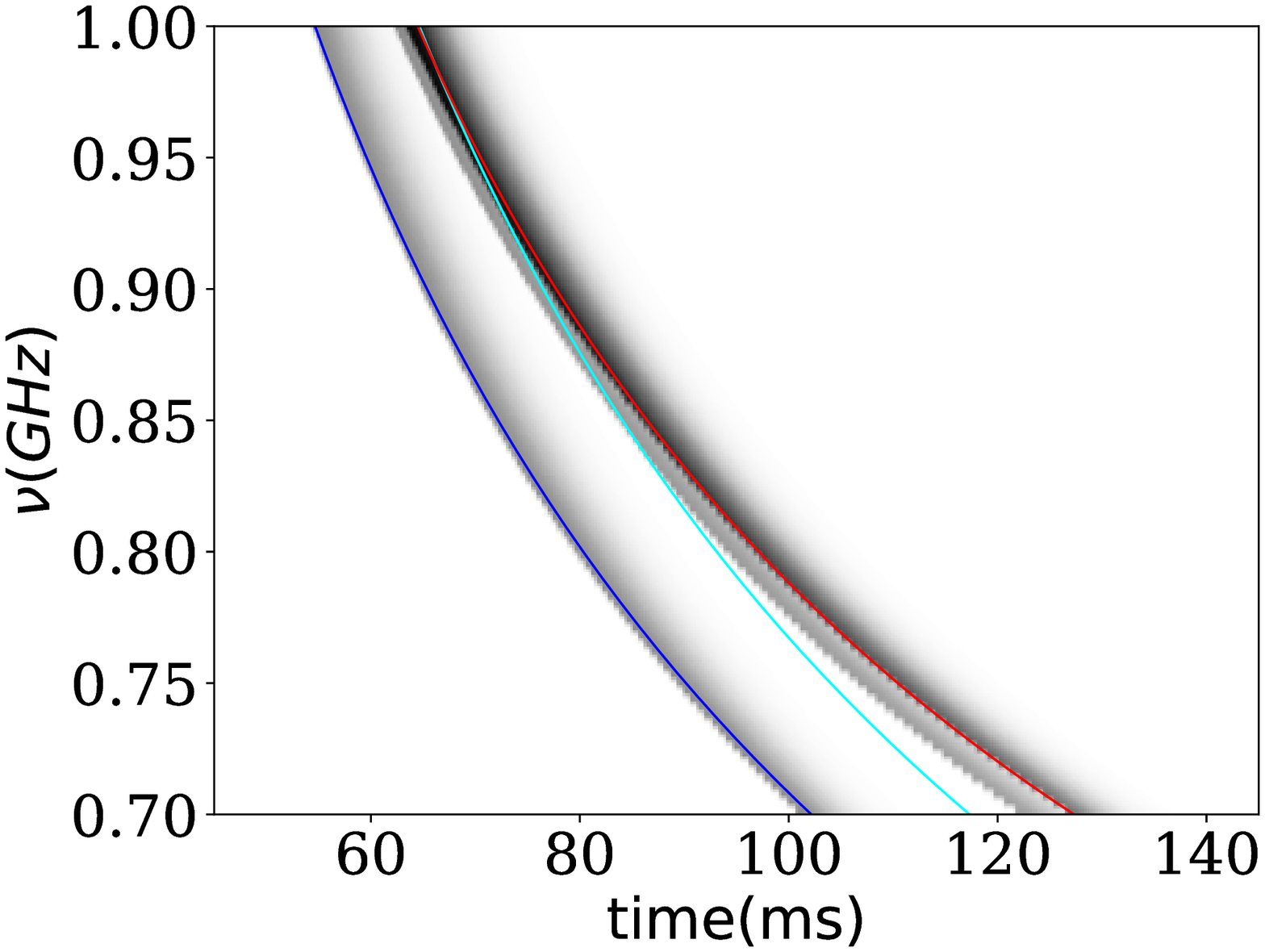}}
  \centerline{\includegraphics[width=8cm]{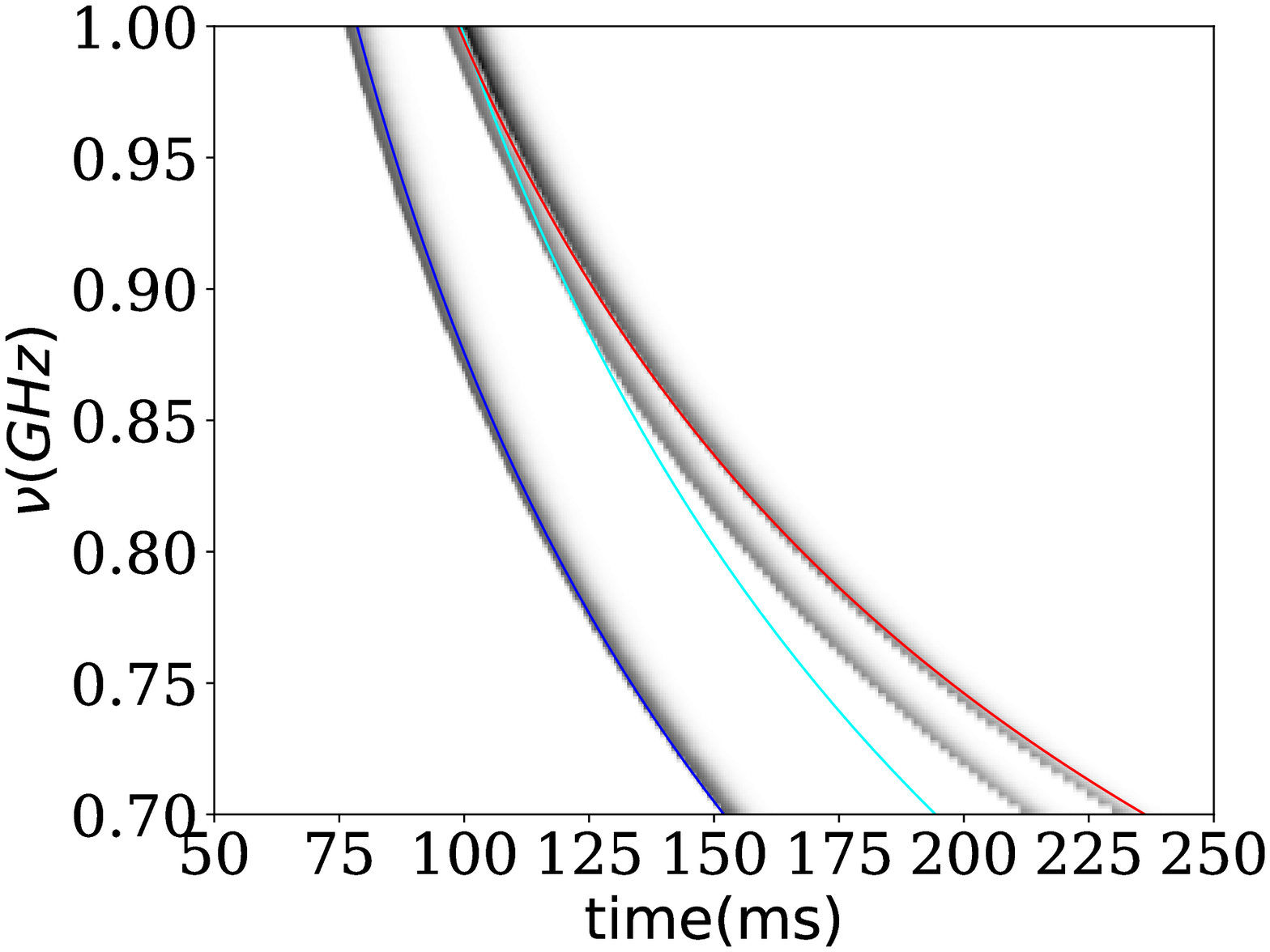}}
  \caption{The simulated radio dispersion signal. In both top and
    bottom panels, from left to right the grey shadow presents the
    time delay signal on frequency for a constant DM($10.7\,$pc
    cm$^{-3}$ in the top panel and $17.3\,$pc\,cm$^{-3}$ in the bottom
    panel), the dispersive delay by the plasma lens, and the total
    delay by the plasma lens. A Gaussian model ($N_0=20\,$pc
    cm$^{-3}$, $\sigma=1000$ AU and located in the Milky Way) is
    adopted for the lens in top panel, and a SPL lens ($h=2$,
    $n_0=0.1$\,cm$^{-3}$, $R_0=10^5$ AU) is adopted in the bottom
    panel. The blue, cyan and red curve show the analytical curves of
    Eqs.\ref{eq:analdm} and \ref{eq:analdm2}. See the text for more
    details. }
  \label{fig:tdl_dm}
\end{figure}
%
We present an example of plasma lenses in the Milky Way by simulating a
toy radio burst signal with an intrinsic Gaussian model of width $\sigma=5$
milliseconds. We assume that there are no intrinsic delays between
frequencies, and propagate it through a clump of plasma. Three
different cases of plasma are studied: in the first one, we assume
that the plasma is uniformly distributed, i.e. the classical case
without lensing. The delay time is calculated from the theoretical
prediction
\be
t(\nu)= 4.15\, {\rm ms}\, \rund{{\rm DM} \over \nu^2},
\elabel{analdm}
\ee
where $\nu$ is given in units of GHz. In the two other cases we place
a plasma lens between us and the source. In the second one, we do not
include the geometric delay, and in the third we include all the delay
effects from plasma lensing. In the top panel, we adopt a Gaussian
lens with $N_0=20\,$pc\,cm$^{-3}$, $\sigma=1000$ AU. The
characteristic radius depends on the frequency. At $1(0.7)$GHz, it is
about $0.12(0.18)$ arcsec. On the source plane, the corresponding
caustic is a circle with radius of about $0.13(0.2)$ arcsec. The radio
source is placed at an angular separation of $0.45$ arcsec on the
source plane, which is far outside the caustics. At this position the
corresponding DM is about $10.7$ pc\,cm$^{-3}$. In
Fig.\,\ref{fig:tdl_dm}, the gray shadows show the three cases from
left to right in order: the constant DM, a lens without geometric
delay, and the total delay. As a guide for comparison, the blue line
represents Eq.\,\ref{eq:analdm} for a constant
DM$=11\,$pc\,cm$^{-3}$. We plot the total delay with another
  constant DM$=12.2\,$pc\,cm$^{-3}$ shown by the cyan line, which nearly
  overlaps the total simulated grey shadow.  As we can see the
geometric effect changes the slope of the curve. Such a relationship
can be represented by taking into account the higher order
effects, as we know that the geometric delay is proportional to
$\lambda^4$,
\be
t(\nu)= 4.15\, {\rm ms}\, \rund{{\rm DM} \over \nu^2} + b \rund{{\rm DM}^2 \over \nu^4},
\elabel{analdm2}
\ee
where the same DM is used, and $b$ is a free parameter determined by
the geometric effect of the lens. The red line is another fit
using Eq.\,\ref{eq:analdm2} with DM$=11\,$pc\,cm$^{-3}$ and $b=0.04$
ms.  Such a fit is different from other empirical relations,
for example, Faraday rotation measure caused by the magnetic field
\citep[e.g.][]{frbreview}.
In the other example (bottom panel of Fig.\,\ref{fig:tdl_dm}), we use
a SPL lens with $h=2$, $n_0=0.1$\,cm$^{-3}$, $R_0=10^5$ AU and
$\theta_C=0.005R_0$, which gives $\theta_0\approx0.17$ arcsec, and
$0.21$ arcsec of caustics on the source plane at frequency of 1
GHz. At $0.7$ GHz, the two scales are 0.21 arcsec and 0.29 arcsec
respectively. The radio source is separated from the lens by $0.25$ arcsec
on the source plane, still out of the caustics. The corresponding
DM$=17.3\,$pc\,cm$^{-3}$. To plot the three analytical curves, we use
DM$=17\,$pc\,cm$^{-3}$ for blue line, DM$=22\,$pc\,cm$^{-3}$ for the cyan
line, and DM$=17\,$pc\,cm$^{-3}$, $b=0.07$ for the red line.

\section{Discussion}

Plasma lensing can cause a significant frequency-dependent time delay
effect.  The geometric contribution to the time delay may provide a
large fraction of the total delay, which strongly depends on the
relative distance and properties of the lens.

We found that when the lens is located at the mid-point between the
source and observer, refractive lensing can easily cause large time
delays, sometimes even larger than the dispersive term. However, such
events only happen when the lens perfectly aligns with the source and
the density gradient is large. The realistic probability of such a
case occuring is low (assuming the cross-section is a few
milli-arcsec$^2$). On the other hand, when the lens is located within
the Milky Way, it also causes a non-negligible time delay (larger than
dispersive delay in all the cases that we presented in this work). We
expect the chance of such an event may be higher, and evidence for
such refractive effects already exists in some observations of radio
sources. Since they have different wavelength dependence, some of the
abnormal DMs found in FRBs may be caused by the geometric effects of
plasma lensing.

In addition, the frequency-dependent magnification can also provide strong constraints to lensing modelling and help us obtain a better estimate for the electron density as well as the intrinsic properties of the source.

In our study, we adopt the thin lens approximation. As discussed, due
to the geometric effect, the radio signal at different frequency
experiences different DM, thus the accuracy of the thin lens
approximation may not be sufficient when the small scale variation of
the plasma is strong, and is worth further study. Such an
approximation is widely used in the gravitational lensing
community. However, the distribution of ionized gas is more
complicated than dark matter halos due to turbulence and related
dynamical phenomena. The thin lens approximation with a single lens
plane may not be sufficient for such studies. Multi-plane lens and
more complex diffuse distribution models are necessry in future work.

Since FRBs have a large event rate, $10^3-10^4$ per day all sky
\citep[e.g.][]{frbreview}, it is expected that an FRB may pass through
a foreground object to reach an observer \citep{pro19,fed19}. In this
case, the dispersion of the FRBs will deviate from the classic
dispersion relation, especially at low frequency. Therefore, a
detailed analysis of the FRB dispersion relation would be helpful to
study the properties of plasma along the line of sight. Insights into
plasma lenses are important because they are difficult to study at all
distance scales. Besides lenses in intervening galaxies, even plasma
lenses that are near to the observer in the ISM are difficult to
detect in general. With knowledge of the detailed dispersion
properties of FRBs, one can study the properties of near-source
plasma, e.g., the inhomogeneous properties of supernova remnants,
pulsar wind nebulae, and HII regions.

\section*{Acknowledgments}
We thank the referee for very valuable and detailed constructive
comments to the manuscript. We also thank Jenny Wagner, Bing Zhang,
Artem Tuntsov, Guoliang Li for interesting discussions and helpful
comments on the draft. XE is support by NSFC Grant No. 11873006.

\bibliography{eplasma}

\end{document}